\documentclass[twocolumn,twocolappendix]{aastex631}

\providecommand{\sorthelp}[1]{}
 
\usepackage{graphicx}
\usepackage{amssymb}
\usepackage{epstopdf}
\usepackage{graphicx}
\usepackage{bm}
\usepackage{xcolor} 
\usepackage{soul}
\usepackage[caption=false]{subfig}
\usepackage[version=4]{mhchem}

\newcommand{\be}{\begin{eqnarray}}

\newcommand{\ee}{\end{eqnarray}}

\usepackage[normalem]{ulem}

\providecommand{\sorthelp}[1]{}
 
\begin{document}

\preprint{APS/123-Q ED}

\title{Quijote-PNG: Simulations of primordial non-Gaussianity and the information content of the matter field power spectrum and bispectrum}

\newcommand{\ens}{Laboratoire de Physique de l'\'{E}cole Normale Sup\'{e}rieure, ENS, Universite PSL, CNRS, Sorbonne Universit\'{e}, Universit\'{e} de Paris, F-75005 Paris, France}
\newcommand{\cnrs}{Sorbonne Universit\'{e}, CNRS, UMR 7095, Institut d'Astrophysique de Paris, 98 bis bd Arago, 75014 Paris, France}
\newcommand{\cca}{Center for Computational Astrophysics, Flatiron Institute, 162 5th Avenue, New York, NY 10010, USA}
\newcommand{\bologna}{Dipartimento di Fisica e Astronomia, Alma Mater Studiorum - University of Bologna, Via Piero Gobetti 93/2, 40129 Bologna BO, Italy}
\newcommand{\inaf}{INAF - Osservatorio Astronomico di Bologna, Via Piero Gobetti 93/3, 40129 Bologna BO, Italy}
\newcommand{\infn}{INFN - Istituto Nazionale di Fisica Nucleare, Sezione di Bologna, Viale Berti Pichat 6/2, 40127 Bologna BO, Italy
}
\newcommand{\infnPad}{INFN, Sezione di Padova, via Marzolo 8, I-35131, Padova, Italy}
\newcommand{\mpa}{Max-Planck-Institut f\"ur Astrophysik, Karl-Schwarzschild-Straße 1, 85748 Garching, Germany}
\newcommand{\ICC}{ICC, University of Barcelona, IEEC-UB, Martí i Franquès, 1, E-08028 Barcelona, Spain}
\newcommand{\ICREA}{ICREA, Pg. Lluís Companys 23, Barcelona, E-08010, Spain}
\newcommand{\Galilei}{Dipartimento di Fisica e Astronomia “G. Galilei”,Università degli Studi di Padova, via Marzolo 8, I-35131, Padova, Italy}
\newcommand{\uwc}{Department of Physics and Astronomy, University of the Western Cape, Cape Town 7535, South Africa}
\newcommand{\princeton}{Department of Astrophysical Sciences, Princeton University, 4 Ivy Lane, Princeton, NJ 08544 USA}
\author{William R Coulton}
\affiliation{\cca}
\author{ Francisco Villaescusa-Navarro}
\affiliation{\cca}
\affiliation{\princeton}
\author{Drew Jamieson}
\affiliation{\mpa}
\author{Marco Baldi}
\affiliation{\bologna}
\affiliation{\inaf}
\affiliation{\infn}
\author{Gabriel Jung}
\affiliation{\Galilei}
\affiliation{\infnPad}
\author{Dionysios Karagiannis}
\affiliation{\uwc}
\author{Michele Liguori}
\affiliation{\Galilei}
\affiliation{\infnPad}
\author{Licia Verde}
\affiliation{\ICREA} 
\affiliation{\ICC}
\author{Benjamin D. Wandelt}
\affiliation{\cnrs}
\affiliation{\cca}

\date{\today}
\begin{abstract}
Primordial non-Gaussianity (PNG) is one of the most powerful probes of the early Universe and measurements of the large scale structure of the Universe have the potential to transform our understanding of this area. However relating measurements of the late time Universe to the primordial perturbations is challenging due to the non-linear processes that govern the evolution of the Universe. To help address this issue we release a large suite of N-body simulations containing four types of PNG: \textsc{quijote-png}. These simulations were designed to augment the \textsc{quijote} suite of simulations that explored the impact of various cosmological parameters on large scale structure observables. Using these simulations we investigate how much information on PNG can be extracted by extending power spectrum and bispectrum measurements beyond the perturbative regime  at $z=0.0$. This is the first joint analysis of the PNG and cosmological information content accessible with power spectrum and bispectrum measurements of the non-linear scales. We find that the constraining power improves significantly up to $k_\mathrm{max}\approx 0.3 h/{\rm Mpc}$, with diminishing returns beyond as the statistical probes signal-to-noise ratios saturate. This saturation emphasizes the importance of accurately modelling  all the contributions to the covariance matrix. Further we find that combining the two probes is a powerful method of breaking the degeneracies with the $\Lambda$CDM parameters. 
\end{abstract}

\pacs{Valid PACS appear here}

\section{Introduction}
\label{sec:intro}
Measurements of the cosmic microwave background (CMB) \citep{fixsen1997,bennett2012,planck2016-l01} and of the large scale structure (LSS) \citep{Alam_2017,dAmico_2020,Ivanov_2020,Kobayashi_2021} have revolutionized our understanding of the primordial Universe, demonstrating that it was nearly homogeneous and isotropic with small, almost scale-invariant perturbations. However, wide classes of theoretical models, ranging from inflationary to ekpyrotic models \citep{Lehners_2010,Martin_2016,Meerburg_2019}, can explain these observations. There are two proposed observational channels that have great power at discriminating between these theories: measurements of primordial gravitational waves and measurements of primordial non-Gaussianity (PNG) \citep[see e.g][for a recent overview of prospects for learning about inflation]{Achucarro_2022}.

Primordial non-Gaussianity is interesting due to its strong sensitivity to the early universe physics: PNG can reveal information about the field content of the Universe, the strength of interactions and particle content \citep{Chen_2007}. Work by \citet{Green_2020} suggests that PNG measurements offer a method to verify the quantum nature of the primordial perturbations.  The information encoded in PNG is complementary to other probes, such as primordial gravitational waves and the slope and running of the primordial power spectrum, that contain information about the duration and energy scale of inflation \citep[see e.g.][for a review]{Lyth_1999}.

Given the strong theoretical motivation, there have been many searches for primordial non-Gaussianity \citep{Ferreira_1998,Komatsu_2002,komatsu2003,Creminelli_2006,Slosar_2008,Leistedt_2014}, with the leading constraints from measurements of the CMB by the \textit{Planck} satellite \citep{planck2016-l09}. The CMB is an ideal probe for studying PNG as CMB anisotropies are linearly related to the primordial perturbations and so CMB measurements can be straightforwardly related to the primordial universe.  Whilst upcoming CMB experiments will improve upon existing constraints \citep{SO_2019,S4_2016}, the challenges from foregrounds and the limitations imposed by silk-damping and the 2D nature of the CMB mean that improvements needed to reach the most interesting theoretical levels will be highly challenging \citep{Hill_2018,Jung_2018,Coulton_2020,Kalaja_2021,Coulton_2022c}. 
 
 Constraints on PNG from measurements of the LSS promise, in principle, to far surpass existing constraints \citep{Meerburg_2017,Slosar_2019}. However, to date, the important measurements of PNG from LSS \citep{Slosar_2008,Leistedt_2014,Giannantonio_2014,Ross_2013,Ho_2015,Castorina_2019,Mueller_2021,Cabass_2022a,Cabass_2022b,DAmico_2022} have not reached the level of the CMB. 
 
 This is due to complexities of large scale foregrounds \citep{Pullen_2013,Rezaie_2021}, understanding the optimal statistics to use, and the challenge of modelling the LSS. Unlike the CMB, measurements of the late-time LSS are non-linearly related to the primordial perturbations which complicates inferences about the primordial universe.  For the largest scales, powerful perturbation theory models have been developed and recently applied to data \citep{Baldauf_2011,Cabass_2022a,Cabass_2022b,DAmico_2022}. However these methods will not be able to probe arbitrarily small scales, and thus it is unclear whether alternative approaches may be better suited to the problem. This is particularly interesting as recent work has suggested that information of PNG could be separated out from other small scale effects by exploiting the locality of gravitational and baryonic processes \citep{Baumann_2021}. 
 
 To aid our understanding of the relation between the primordial non-Gaussianity and late time observables, we have run an extensive suite of simulations with PNG. This work builds on the results of numerous previous investigations: \citet{Dalal_2008,Desjacques_2009} generated simulations containing \emph{local} PNG and used these to discover and understand the impact of PNG on scale dependent bias. This work was followed by \citet{Wagner_2010,Scoccimarro_2012} who developed and implemented methods to generate simulations with PNG beyond the \emph{local}. In this work we use the methods developed in \citet{Scoccimarro_2012} to generate a large ensemble of simulations with four different PNG shapes. 
 
 These simulations were designed to fit within the \textsc{quijote} suite of simulations \citep{Villaescusa-Navarro_2020}; a large suite of simulations designed to quantify the information content on generic summary statistics and to provide enough training data for machine learning algorithms.
 By making this choice our PNG simulations can be used within a consistent framework to study how uncertainties in the standard cosmological parameters impacts inferences of PNG. For example, whilst previous work has shown that the halo mass function \citep{Lucchin_1988,LoVerde2011}, the matter probability density function \citep{Valageas_2002,Uhlemann_2018,Friedrich_2020}, topological measures \citep{Cole_2020,Biagetti_2022} and field level analyses \citep{Andrews_2022} are potentially powerful probes of PNG, simulations are required to validate theoretical predictions or model these novel probes. An aim of this work is to help facilitate such analyses.

 As both a validation of our simulations and a first application we explore the properties of the matter power spectrum and bispectrum in these simulations. On the largest scales the impact of PNG on the matter power spectrum and bispectrum is well understood \citep[e.g.][]{Desjacques_2009,Baldauf_2011,Karagiannis_2018} and so these measurements can be used to validate our simulations. However our simulations also allow us to push beyond the perturbative regime ($k \approx 0.1~h/{\rm Mpc}$ at $z=0$) and explore the information content available on smaller scales. Recent work by \citet{Hahn_2020} has shown that for other cosmological parameters, small scale bispectrum measurements of the halo and galaxy field potentially contain large amounts of information. Through the use of a simulation-based covariance matrix we are able to assess the importance of all the contributions, including the super-sample covariance, for parameter constraints.

This paper and our companion paper, \citet{Jung_2022}, are the first two papers in a series dedicated to investigating how we can learn more about primordial non-Gaussianity from upcoming measurements of the LSS.
 
In our companion paper, \citet{Jung_2022}, we explore the properties of matter power spectrum and bispectrum from a different and complementary perspective: we use the modal bispectrum \citep{Fergusson2012,Schmittfull2013} to measure the matter bispectrum and then develop and validate a set of optimal compressed statistics, which enable the information in the matter power spectrum and bispectrum to be captured within a data vector of length the number of parameters. Additionally the \citet{Jung_2022} analysis focuses on measurements at $z=1.0$, the redshift range relevant for upcoming surveys such as Euclid \citep{Amendola_2018}, whereas here we examine $z=0.0$. Through this complementarity we can better understand the importance of small scale information across redshifts. We stress that both of these works consider an idealized setup - the 3D matter field is not directly observable and we neglect all observational effects and their associated challenges, which will likely limit future analyses - and thus it is difficult to directly relate constraints reported here to future surveys. Further the unique scale dependent bias feature that PNG introduces to biased tracers, means that the measurements of biased tracers could contain more information than the matter field, on the same scales. However the aim of these works is not to generate specific predictions but rather assess the value of pushing to smaller scales and provide simulations that can be used as a sandbox to develop intuition into new estimators in an simplified environment. This represents a first step, and motivation, towards more complex and realistic analyses. 
 
 These two papers represent the first comprehensive investigation into what can be learnt about PNG from small scale measurements of the matter bispectrum and power spectrum. These papers are accompanied by the release the simulation data and our power spectrum and bispectrum measurements. In the next papers we will consider the information content in the halos \citep{Coulton_2022a} and the information accessible from analyses of the field.
 
 Our paper is structured as follows: in Section \ref{sec:PNG_intro} we briefly overview the shapes of non-Gaussianity considered in this work and in Section \ref{sec:simulations} we describe our simulations of PNG. We describe our power spectrum and bispectrum estimators in Section \ref{sec:statistics} and then present our measurements in Section \ref{sec:validation_ICs}. In Section \ref{sec:FisherMethods} we describe the details of how we compute our Fisher forecasts and then explore the constraining power of the bispectrum and power spectrum measurements in Section \ref{sec:cosmo_constraints}. We then summarize our key results in Section \ref{sec:conclusions}. In the three appendices we describe the details of the generation of non-Gaussian initial conditions (Appendix \ref{app:ic_kernels}), the details of the computation of the power spectrum and bispectrum covariance matrices (Appendix \ref{app:covarianceMatrix}), joints constraints of PNG and $\Lambda$CDM extensions and finally the impact of a prior on the power spectrum convergence (Appendix \ref{app:priorPk}).

\section{Primordial non-Gaussianity} \label{sec:PNG_intro}
Whilst primordial non-Gaussianity refers to any deviation from Gaussian initial conditions, in this work we restrict our analysis to the primordial bispectra, $B_{\Phi}(k_1,k_2,k_3)$, defined as
\begin{align}
    &\langle \Phi(\mathbf{k}_1) \Phi(\mathbf{k}_2) \Phi(\mathbf{k}_3) \rangle \nonumber \\ &=  (2\pi)^3 \delta^{(3)}(\mathbf{k}_1+\mathbf{k}_2+\mathbf{k}_3)B_{\Phi}(k_1,k_2,k_3).
\end{align}
where $\Phi(\mathbf{k})$ is the primordial potential. We further restrict our analysis to four different primordial bispectra shapes, detailed below. These shapes 
are studied as they accurately approximate theoretically well motivated shapes, provide insight into the primordial physics and can be generated by a range of generic methods \citep[see e.g.][for reviews of PNG]{Chen_2010,Achucarro_2022}.

The first shape we consider is the \emph{local} shape and has the primordial bispectrum
\begin{align}
    B^{\mathrm{local}}_{\Phi}(k_1,k_2,k_3) = & 2 f_{\mathrm{NL}}^{\mathrm{local}} P_\Phi(k_1)P_\Phi(k_2)+  \text{ 2 perm.}
\end{align}
where $P_\Phi(k_1)$ is the primordial power spectrum and $f_{\mathrm{NL}}^{\mathrm{local}}$ is the amplitude of this type of non-Gaussianity. \emph{Local} non-Gaussianity is of observational interest as it is a powerful probe of the primordial field content. \citet{Maldacena_2003,Creminelli_2004} showed that in single-field slow roll inflation, the amplitude of the \emph{local} bispectrum is $<O(\eta,\epsilon)$, where $\epsilon$ and $\eta$ are the slow roll parameters, thus a measurement of larger \emph{local} non-Gaussianity would present difficulties for slow-roll, single-field inflation. On the other hand, observable levels of \emph{local} non-Gaussianity can be generated in multi-field inflationary models, such as the curvaton or modulated reheating models \citep{Lyth_2002,Dvali_2004}, and alternatives to inflation models \citep{Lehners_2010}. Note that whilst there is on-going discussion \citep{Pajer_2013,Matarrese_2021} as to whether the squeezed-limit bispectrum exactly vanishes for single-field models, this does not impact the power of a detection of \emph{local} $>>O(\epsilon,\eta)$ to rule out these models.

The second shape we consider is the \emph{equilateral} non-Gaussianity, which has the following primordial bispectrum
\begin{align}
  &   B^{\mathrm{equil.}}_{\Phi}(k_1,k_2,k_3) = 6 f_{\mathrm{NL}}^{\mathrm{equil.}}\Big[- P_\Phi(k_1)P_\Phi(k_2)+\text{ 2 perm.} \nonumber \\ &  
  -2 \left( P_\Phi(k_1)P_\Phi(k_2)P_\Phi(k_3) \right)^{\frac{2}{3}} +  P_\Phi(k_1)^{\frac{1}{3}}P_\Phi(k_2)^{\frac{2}{3}}P_\Phi(k_3)  \nonumber \\ & + \text{5 perm.}\Big]
\end{align}
\emph{Equilateral} non-Gaussianity well approximates non-Gaussianities generated in Dirac-Born-Infeld inflation \citep{Alishahiha_2004}, ghost inflation \citep{Arkani-Hamed_2004} and generically in models with local, derivative interactions \citep{Cheung_2008,Cheung_2008b}.

The final shape that we consider is the \emph{orthogonal} shape. The \emph{orthogonal} bispectrum, along with the \emph{equilateral} bispectrum, is used to span the types of non-Gaussianity within the effective field theory of inflation (EFTi)\citep{Cheung_2008,Cheung_2008b}. Measurements of the \emph{equilateral} and \emph{orthogonal} bispectra can be used to constrain the parameters of EFTi, including constraining the sound speed of primordial perturbations \citep{Senatore_2010,planck2013-p09a}. The full shape of the \emph{orthogonal} bispectrum in EFTi is not separable (i.e expressible as $F(k_1)G(k_2)H(k_3) +\text{perms.}$) and so is challenging to simulate and analyze. Two approximations have been developed in \citet{Senatore_2010}; the first approximation was performed to approximate the \emph{orthogonal} bispectrum in the CMB and hence we refer to this as \emph{orthogonal-CMB}. It has the bispectrum
\begin{align}
    B^{\mathrm{ortho-CMB}}_\Phi&(k_1,k_2,k_3) = 6 f_{\mathrm{NL}}^{\mathrm{ortho-CMB}}\Big[-3 P_\Phi(k_1)P_\Phi(k_2) \nonumber \\ &  
   +\text{ 2 perm.}  -8 \left( P_\Phi(k_1)P_\Phi(k_2)P_\Phi(k_3) \right)^{\frac{2}{3}} +  \nonumber \\ & 3P_\Phi(k_1)^{\frac{1}{3}}P_\Phi(k_2)^{\frac{2}{3}}P_\Phi(k_3)  + \text{5 perm.}\Big].
\end{align}
Whilst this functional form is a good approximation for the CMB, which is a 2D projection of the primordial bispectrum, it is less accurate for measurements of the LSS \citep{Senatore_2010,Creminelli_2011}. In particular, its squeezed-limit divergence, which is suppressed for the 2D  CMB, differs from the EFTi \emph{orthogonal} shape. This different squeezed limit has important consequences for scale dependent bias measurements and for the detectibility. Despite these issues, we include it in our suite of simulations as the resulting scale-dependent bias ($\sim1/k$) is different from the \emph{local} case ($\sim1/k^2$), and \emph{orthogonal-LSS} (see below) and \emph{equilateral} shapes ($\sim1$), so the simulations help span possible PNG signatures \citep{Schmidt_2010}. Further \emph{folded} bispectra, which can arise if the assumption of the Bunch-Davis vacuum is violated, is well approximated by a linear sum of this shape with the \emph{equilateral} shape \citep{Meerburg_2010,Chen_2007}. The impact of these subtleties on the halo field, where scale dependent bias is important, is further explored in \citet{Coulton_2022a}.

The second approximation, also developed by \citet{Senatore_2010} and hereafter referred to as \emph{orthogonal-LSS}, has a bispectrum given by
\begin{align} \label{eq:bis_or_lss}
   & B^{\mathrm{ortho-LSS}}_\Phi(k_1,k_2,k_3) = \nonumber \\ & 6 f_{\mathrm{NL}}^{\mathrm{ortho-LSS}}
        \left(P_\Phi(k_1)P_\Phi(k_2)P_\Phi(k_3)\right)^{\frac{2}{3}}\Bigg[ \nonumber \\ &  -\left(1+\frac{9p}{27}\right) \frac{k_3^2}{k_1k_2} + \textrm{2 perms} +\left(1+\frac{15p}{27}\right)  \frac{k_1}{k_3} \nonumber \\ &   + \textrm{5 perms}  -\left(2+\frac{60p}{27}\right)  \nonumber \\ & +\frac{p}{27}\frac{k_1^4}{k_2^2k_3^2} + \textrm{2 perms}  -\frac{20p}{27}\frac{k_1k_2}{k_3^2}+ \textrm{2 perms} \nonumber \\ & -\frac{6p}{27}\frac{k_1^3}{k_2k_3^2} + \textrm{5 perms}+\frac{15p}{27}\frac{k_1^2}{k_3^2} + \textrm{5 perms}
    \Bigg],
\end{align}
where
\begin{align}
    p=\frac{27}{-21+\frac{743}{7(20\pi^2-193)}} \, .
\end{align}
This shape is a better approximation to the EFTi \emph{orthogonal} shape for the matter field and exhibits the correct squeezed limit scaling. 
Note -- as is discussed in Appendix \ref{app:ic_kernels} -- that the \emph{orthogonal-LSS} shape we consider is slightly modified to account for the fact that our simulations are not scale invariant, i.e. $n_s\neq1$.

\section{Simulations}
\label{sec:simulations}

In this work we extend the \textsc{quijote} simulations by running a set of simulations at the \textsc{quijote} fiducial cosmology with primordial non-Gaussianity: \textsc{quijote-png}. These simulations have been run with the same settings (i.e. PM grid size, force resolution...etc) as the original \textsc{quijote} simulations. Additionally, \textsc{quijote-png} uses simulations with matching random seeds to compute partial derivatives. Thus these simulations are ideal for investigating the information in PNG jointly with other cosmological parameters. In Tables \ref{tab:simulation_params} and \ref{tab:png_params} we summarize the properties of the new simulations. In this section we describe the details of the PNG initial conditions and the simulations. We refer the reader to \citet{Villaescusa-Navarro_2020} for full details on the \textsc{quijote} simulations. All the PNG simulations are run with the $\Lambda$CDM cosmological parameters shown in Table \ref{tab:simulation_params} .
 They follow the evolution of $512^3$ dark matter particles from $z=127$ down to $z=0$. The following simulation products are publicly available: particle data at z=0.0,\, 0.5,\,1.0,\,2.0, and $3.0$ and the power spectrum and bispectrum measurements. See \url{https://quijote-simulations.readthedocs.io/en/latest/png.html} for more details.
\begin{table*}
    \centering
    \begin{tabular}{c c c c c c c  c c c c c c}
 $\Omega_m$ & $\Omega_\Lambda$  & $ \Omega_b $ & $ \sigma_8 $  & $h$ & $n_s$ & $ \sum m_\nu $  & $w$ & Box size &$ N_{\mathrm{particles}}^{\frac{1}{3}}$ &\# Realizations & ICs & Mass resolution  \\ 
  & &  &  & & &  (eV)  & & (Mpc/h) & & & &  (M$_\odot$/h) \\
    \hline \hline 
 0.3175 & 0.6825 & 0.049 & 0.834 & 0.6711 & 0.9624 & 0.0  & -1 &  1000 & 512 & 500 & 2LPTPNG & $6.56\times10^{11}$  \\
   \end{tabular}
    \caption{A summary of the key properties of our new simulations. The choice of cosmological parameters, random seeds and simulation settings are otherwise chosen to match those used in \citet{Villaescusa-Navarro_2020} \label{tab:simulation_params}. }
\end{table*}

\begin{table*}
    \centering
    \begin{tabular}{c c c c c }
    Name & $f_\mathrm{NL}^{\mathrm{local}}$ & $f_\mathrm{NL}^{\mathrm{equil.}}$ & $f_\mathrm{NL}^{\mathrm{ortho-CMB}}$& $f_\mathrm{NL}^{\mathrm{ortho-LSS}}$  \\
    \hline
    \hline
    LC\_p & 100 & 0 & 0 & 0   \\
    LC\_m & -100 & 0 & 0 & 0   \\
    EQ\_p & 0 & 100 & 0 & 0   \\
    EQ\_m & 0 & -100 & 0 & 0   \\
    OR\_p & 0 & 0 & 100 & 0   \\
    OR\_m & 0 & 0 & -100 & 0   \\
    OR\_LSS\_p & 0 & 0 & 0 & 100   \\
    OR\_LSS\_m & 0 & 0 & 0 & -100   \\
    \end{tabular}
    \caption{ The name convention and $f_\mathrm{NL}$ parameters used for each simulation \label{tab:png_params} .} 
\end{table*}

\subsection{Non-Gaussian Initial Conditions} \label{sec:ic_generation}
To run non-Gaussian simulations we use the method developed in \citet{Scoccimarro_2012} to generate initial conditions that have primordial non-Gaussianity, and we implement several small changes to the public code released by \citet{Scoccimarro_2012}. Our updated code is available at \url{https://github.com/dsjamieson/2LPTPNG}. Here we briefly review this method and refer the reader to \citet{Scoccimarro_2012} for more details. 

Generating initial conditions with \emph{local} non-Gaussianity is straightforward. First, the modes of a Gaussian primordial potential field, $\Phi(\mathbf{k})$, are generated in Fourier space from the input power spectrum.
This field is then inverse Fourier transformed to real-space. The real-space field is squared, mean subtracted, scaled by the desired amplitude, $f_{\mathrm{NL}}^{\mathrm{local}}$, and then added back to the original potential. The resulting field is real Fourier transformed to obtain the modes of the primordial potential with \emph{local} PnG.

In Fourier space this corresponds to a convolution so that the process can be summarized as
\begin{align}
& \Phi^{\mathrm{local}}(\mathbf{k}) = \Phi(\mathbf{k}) \nonumber \\ & + f_{\mathrm{NL}}^{\mathrm{local}} \int \frac{\mathrm{d}^3 k_1}{(2\pi)^3}\frac{\mathrm{d}^3 k_2}{(2\pi)^3} \Phi(\mathbf{k_1})\Phi(\mathbf{k_2})(2\pi)^3\delta^{(3)}(\mathbf{k}_1+\mathbf{k}_2+\mathbf{k}) \nonumber \\ & -  f_{\mathrm{NL}}^{\mathrm{local}} \int \frac{\mathrm{d}^3 k}{(2\pi)^3}\Phi(\mathbf{k})\Phi^*(\mathbf{k}),
\end{align}
where the second term removes the contribution from the mean, $<\Phi^2>$, which otherwise would contribute to background expansion,

Generating other types of primordial bispectra corresponds to adding a kernel to the convolution
\begin{align} \label{eq:kernelConv}
& \Phi^{\mathrm{X}}(\mathbf{k}) = \Phi(\mathbf{k}) +  f_{\mathrm{NL}}^{\mathrm{NG}} \int \frac{\mathrm{d}^3 k_1}{(2\pi)^3}\frac{\mathrm{d}^3 k_2}{(2\pi)^3} \times \nonumber   \\ &K(\mathbf{k}_1,\mathbf{k}_2,\mathbf{k}) \Phi(\mathbf{k_1})\Phi(\mathbf{k_2})(2\pi)^3\delta^{(3)}(\mathbf{k}_1+\mathbf{k}_2+\mathbf{k}).
\end{align}
Note that we impose $K(\mathbf{k_1}, \mathbf{k_2},0) = 0$ as this automatically accounts for the removal of the mean. 

Crucially, for the \emph{orthogonal-CMB}, \emph{orthogonal-LSS} and \emph{equilateral} shapes the kernels required can be written as a linear combination of separable terms, i.e. they can be written as sums over terms with the form  $ K(\mathbf{k}_1,\mathbf{k}_2,\mathbf{k}) = G(\mathbf{k}_1)H(\mathbf{k}_2)M(\mathbf{k})$. These can thus be generated in a similar manner to the \emph{local} with two modifications: first the Gaussian potential modes are filtered before the inverse Fourier transform, and then again after the final Fourier transform operation. The specific kernels we used and their coefficients are given in Appendix \ref{app:ic_kernels}. 

The initial conditions are generated as follows: we generate the primordial anisotropies from a primordial power law spectrum characterized by the amplitude, $A_s$, and tilt, $n_s$. These anisotropies are generated on a grid of size $1024^3$ to minimize the impact of aliasing effects. We add non-Gaussianity via Eq. \ref{eq:kernelConv}. The perturbations are evolved to redshift $z=0.0$ using linear transfer function, $T(k)$ from CAMB \citep{Lewis_2000}. We then scale back its amplitude to $z=127$ using the traditional growth factor, $D(z)$: $P(k,z=127)=D^2(z=127)/D^2(z=0)P(k,z=0)$ \footnote{Note that in this work the growth factor does not include relativistic species and so also the simulations are run without including radiation in the background.  In \citet{Jung_2022} we ran a suite of simulations including this effect and find that it makes a negligible difference to the observables considered here}. The gridded field at $z=127$ is then used with 2LPT to create the initial displacements of particles for our simulation.  We note that we made a small change to the public code to generate initial conditions with $n_s\ne 1$, and with \emph{orthogonal-LSS} non-Gaussianity.

\subsection{N-body Evolution}
After generating the initial conditions, we follow their gravitational evolution down to $z=0$ using the TreePM code \textsc{Gadget-3} \citep{Gadget}. As stated above, we use the same parameter settings as the original \textsc{quijote} simulations. Halos are identified through the Friends-of-Friends (FoF) algorithm \citep{FoF} with a value of the linking length equal to $b=0.2$.

\subsection{Grid assignment}
To measure our statistics we construct a mesh with $N_\mathrm{side}=512$ voxels for each simulation at the considered redshift. We use the Cloud-in-Cell \citep{Hockney_1981} assignment scheme to distribute the particle positions to the grid. We account for this when computing the power spectra and bispectra by deconvolving the Cloud-in-Cell window function \citep{Jing_2005}. Note that the large $N_\mathrm{side}$ used in our analysis means the effects of the window function and aliasing are negligible on the scales of interest: $k\leq0.5~h/{\rm Mpc}$.  This was verified by comparing our measurements to a set of grids generating using a fourth order interpolation scheme with interlacing \citep[see e.g.][for a detailed discussion of these effects]{Sefusatti_2016}. 

\section{Statistical Probes} \label{sec:statistics}
In this work we examine two statistics of the field: the matter power spectrum and matter bispectrum.
\subsection{Matter power spectrum}
The matter power spectrum is defined as 
\begin{align}
\langle \delta^m(\mathbf{k}) {\delta^m}^*(\mathbf{k'}) \rangle = (2\pi)^3 \delta^{(3)}(\mathbf{k}-\mathbf{k'}) P^{mm}(k).
\end{align}
We estimate the binned power spectrum, $\hat{P}(k_i)$ from simulations by computing 
\begin{align}\label{eq:ps_estimator}
\hat{P}^{mm}(k_i) = \frac{1}{N_i} \sum\limits_{k_{i}-\Delta k/2<|\mathbf{k}|\leq k_i+\Delta k/2} \delta^m(\mathbf{k}){\delta^{m}} ^*(\mathbf{k})
\end{align}
where we sum over all modes that lay within a k-space bin of width $\Delta k$ and $N_i$ is the normalization that effectively counts the number of modes within our bin. We use the public Pylians3 \footnote{\url{https://github.com/franciscovillaescusa/Pylians3}} code with bins of width the fundamental mode, $k_{\mathrm{F}}$, from $k_\mathrm{F}$ to $k_\mathrm{max}=0.5~h/{\rm Mpc}$.

\subsection{Matter bispectrum}
The matter bispectrum is defined as 
\begin{align}
\langle \delta^m(\mathbf{k}_1)& \delta^m(\mathbf{k}_2)  \delta^m(\mathbf{k}_3)  \rangle=  \nonumber \\ & (2\pi)^3 \delta^{(3)}(\mathbf{k_1}+\mathbf{k}_2+\mathbf{k}_3) B(\mathbf{k}_1,\mathbf{k}_2,\mathbf{k}_3).
\end{align}

We estimate the bispectrum using the binned bispectrum estimator, as with the implementation details as in \citet{Foreman_2020}\footnote{Note we do not use nbodykit but a second code that was cross validated during the work of \citet{Foreman_2020}} and uses the methods developed in \citep{Scoccimarro_2000,Tomlinson_2019}
The binned bispectrum estimator computes 
\begin{align} 
&\hat{B}(k_i,k_j,k_k) = \frac{1}{N_{ijk} } \sum\limits_{k_{i-1}<|\mathbf{k_1}|\leq k_i}\sum\limits_{k_{j-1}<|\mathbf{k_2}|\leq k_j}\sum\limits_{k_{k-1}<|\mathbf{k_3}|\leq k_k} \nonumber \\ & \delta^h(\mathbf{k}_1)\delta^h(\mathbf{k}_2)\delta^h(\mathbf{k}_3) (2\pi)^3\delta^{(3)}(\mathbf{k}_1+\mathbf{k}_2+\mathbf{k}_3),
\end{align}
which sums over all bispectra configurations that have wavenumbers that fall within three bins and is normalized by $N_{ijk}$, which counts the number of configurations in each bin. Note that as we sum over the discretized modes on the grid, the Dirac Delta function is really a Kronecker Delta for each of the three directions. The estimator is efficiently implemented by Kronecker delta
\begin{align}
 (2\pi)^3\delta^{(3)}(\mathbf{k}_1 + \mathbf{k}_2+\mathbf{k}_3 )  = V\sum\limits_{\mathbf{x}} e^{i \mathbf{x} \cdot (\mathbf{k}_1+\mathbf{k}_2+\mathbf{k}_3)},
\end{align}
where the summation is over the grid points with a volume factor, $V$, and then exchanging the order of the summations. The resulting expression can be rapidly evaluated by fast Fourier transforms. In this work we use the same bins as in \citet{Hahn_2020}: we use bins of width $3 k_\mathrm{F}$ starting from half the fundamental to $k_\mathrm{max}=0.5~h/{\rm Mpc}$.
\begin{figure*}
    \centering
    \subfloat[Primordial power spectrum ]{\label{fig:pk_primordal}
  \includegraphics[width=0.45\textwidth]{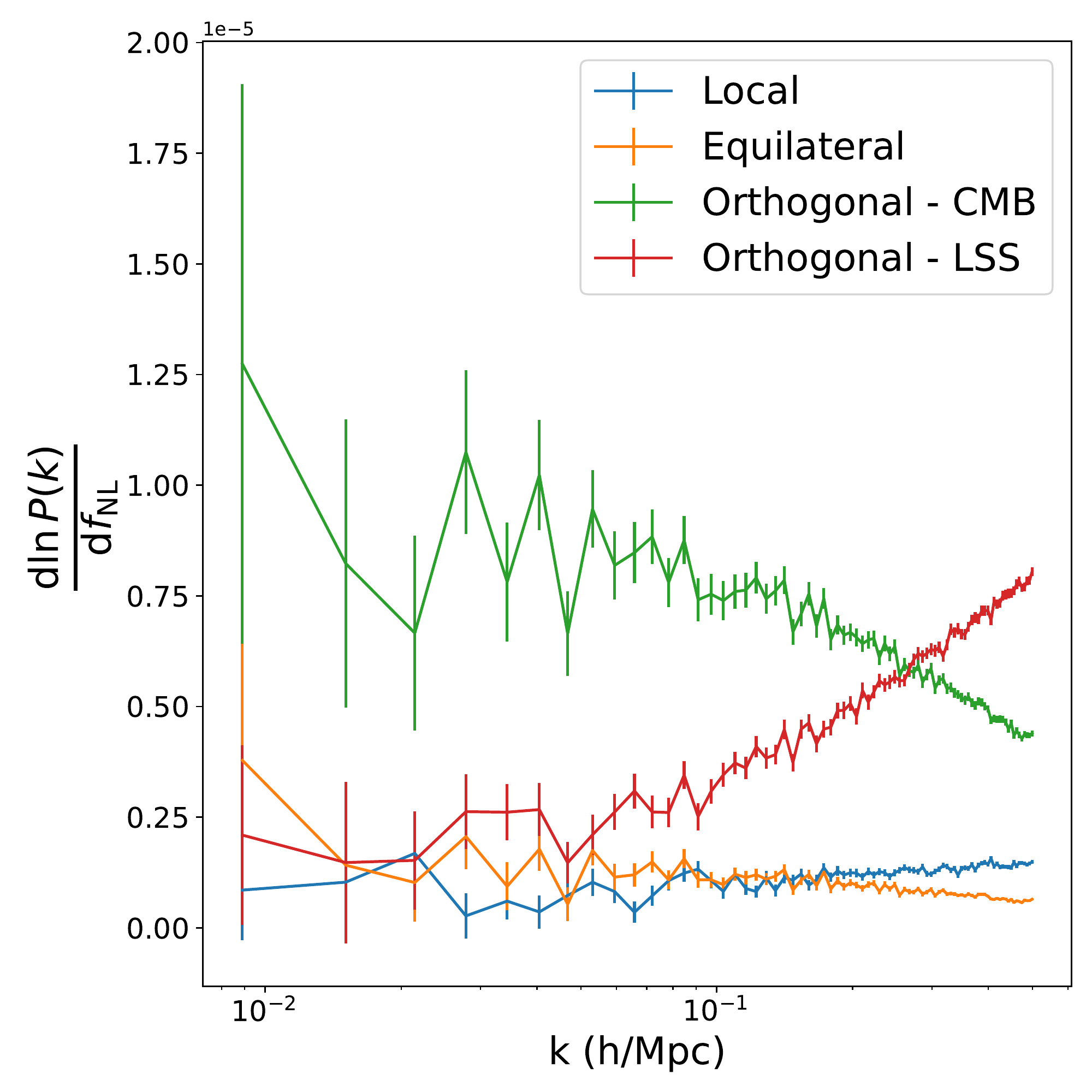}
}
    \subfloat[Matter power spectrum at $z=0.0$ ]{\label{fig:pk_z0}
  \includegraphics[width=0.45\textwidth]{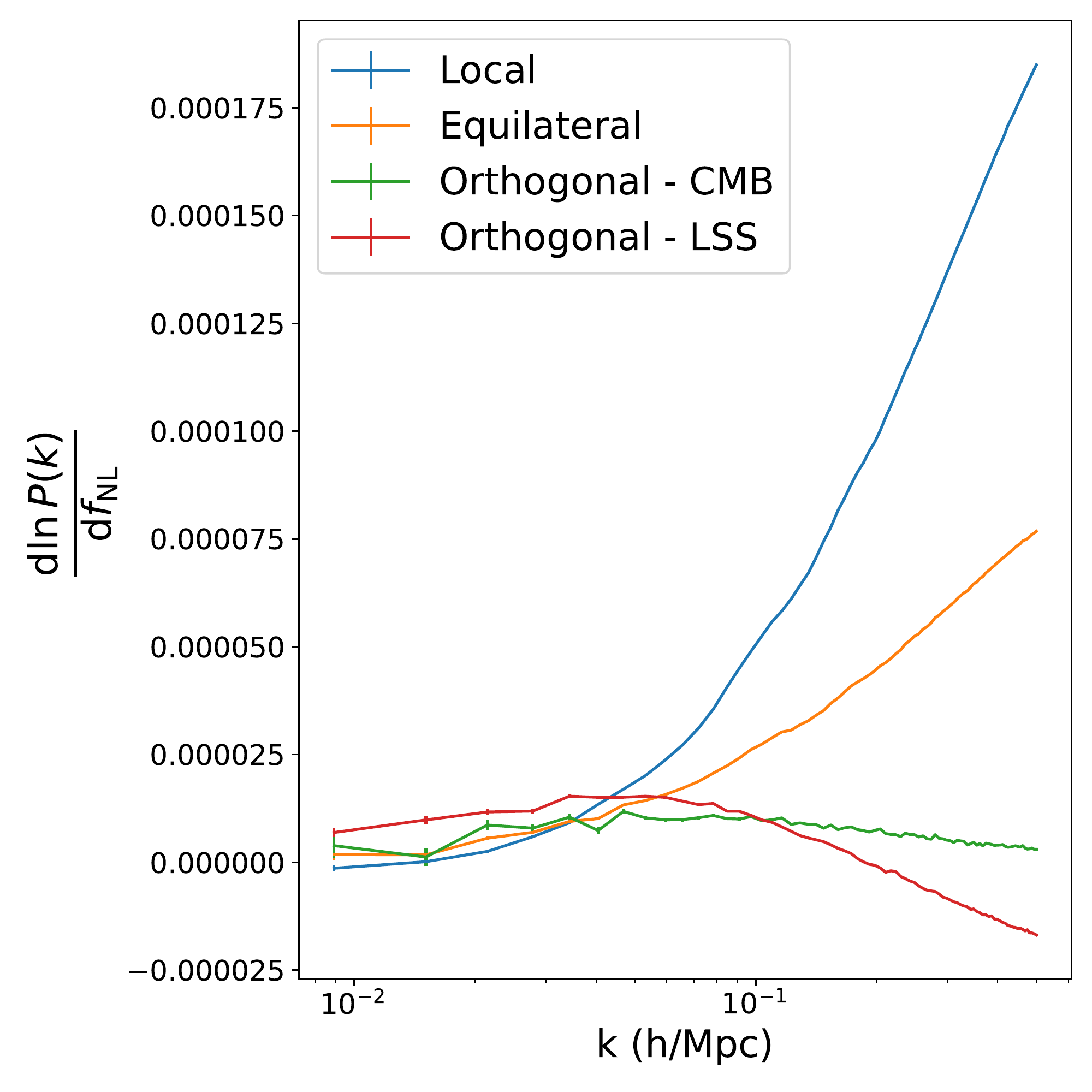}
}   
    \caption{The derivative of the power spectrum with respect to the four different shapes of primordial non-Gaussianity. In Fig. \ref{fig:pk_primordal} we plot the derivatives of the initial conditions whereas in Fig. \ref{fig:pk_z0} shows the derivatives obtained from the simulations at $z=0.0$. The error bars denote the error on the mean. The thick lines show the 1-loop EFT prediction. }
\end{figure*}
\begin{figure*}
    \subfloat[Primordial potential bispectrum]{ \label{fig:bi_primordal}\includegraphics[width=0.45\textwidth]{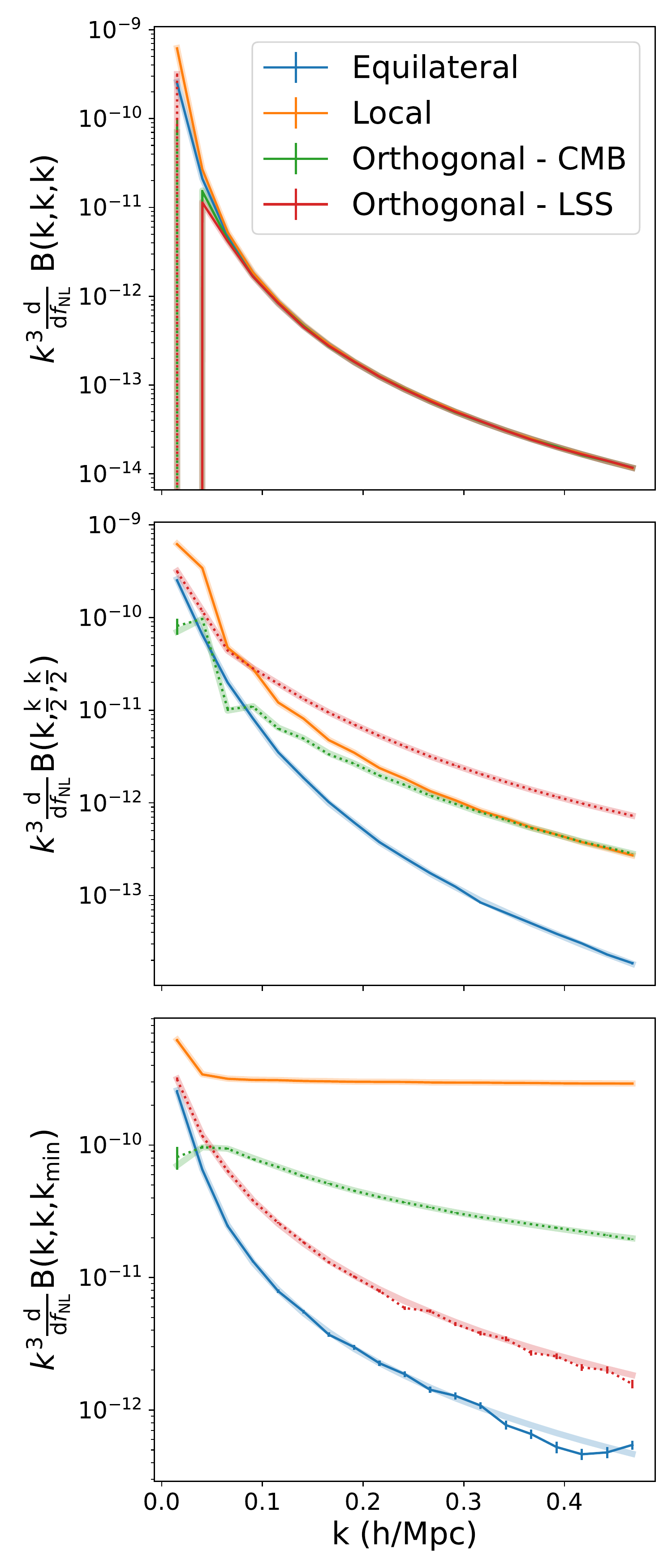}
    }
        \subfloat[Matter bispectrum at $z=0.0$ ]{ \label{fig:bi_z0} \includegraphics[width=0.45\textwidth]{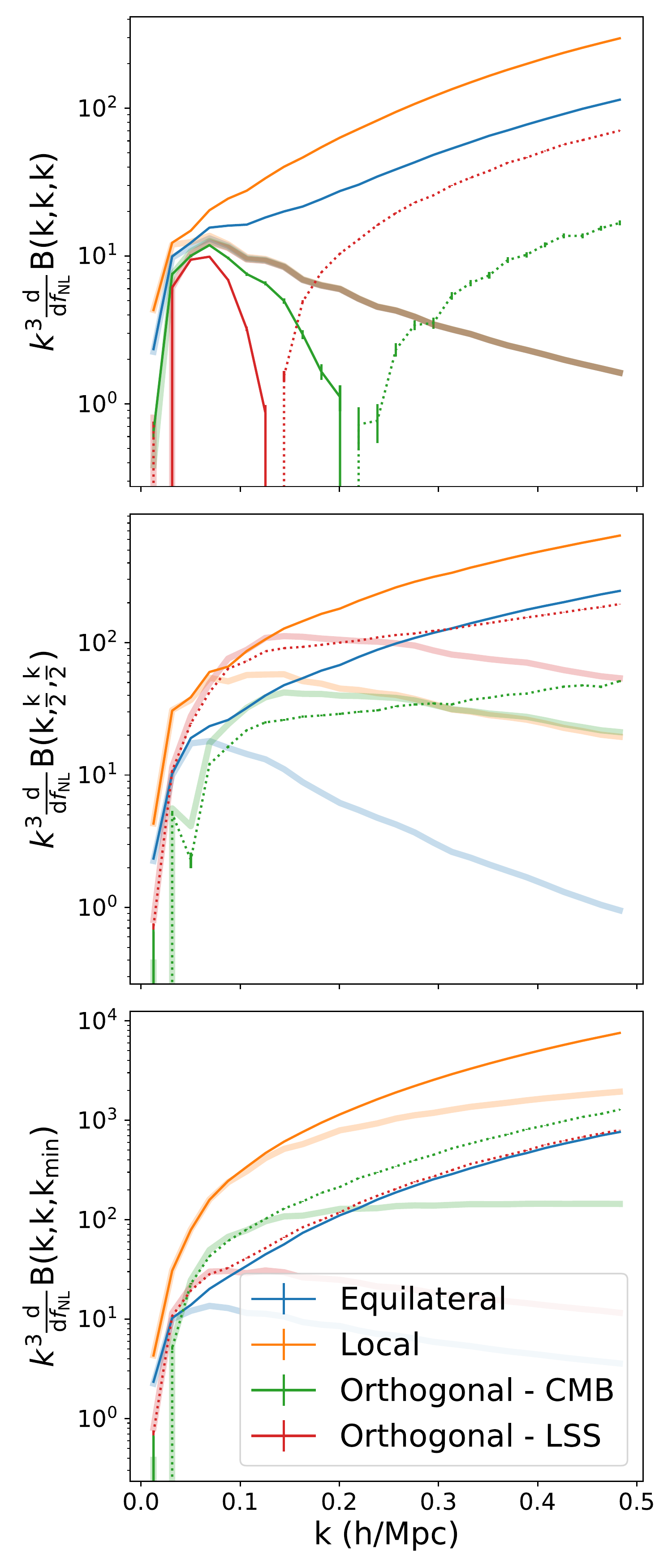}
    }
    \caption{The equilateral (top), folded (middle) and squeezed bispectrum slices of the bispectrum derivative with respect to the four different shapes of primordial non-Gaussianity. In Fig. \ref{fig:bi_primordal} we plot the derivatives of the initial conditions whereas in Fig. \ref{fig:bi_z0} shows the derivatives obtained from the simulations at $z=0.0$. The dotted lines denote regions where the bispectrum is negative, the error bars are the error on the mean of our simulations and the thick shaded line is the tree-level bispectrum prediction  (note the thickness of the theory curve was chosen solely to aid the visualization of overlapping lines). } 
\end{figure*}
\section{Power spectrum and bispectrum measurements}\label{sec:validation_ICs}

We start by exploring the initial conditions used in our simulations. To do so we generate 500 realizations of the Gaussian primordial potential field and 500 realizations for each of the primordial bispectra. In Fig. \ref{fig:pk_primordal} we show the impact of PNG on the primordial potential power spectrum. As is discussed in \citet{Wagner_2010,Scoccimarro_2012}, care needs to be taken that the kernels to generate the bispectrum do not generate unphysical corrections to the power spectrum. We see that there are no divergent corrections to the power spectrum introduced by our method. We do see that there are small corrections, $1\times 10^{-5}f_\mathrm{NL}$, to the power spectrum for the \emph{orthogonal-CMB} non-Gaussianity. These corrections arise due to the loop correction to the power spectrum, some of which give $k$-dependent corrections and others are equivalent to an amplitude renormalization \citep[see the Appendix of][for a detailed discussion]{Wagner_2010}. In principle the scale dependent contributions could be further minimized by different kernel choices, however there is limited theoretical motivation to chose one parameterization over another and hence we opted for the simplest implementation.

In Fig. \ref{fig:bi_primordal} we plot three slices showing how the bispectrum of the primordial initial conditions is affected by PNG. In addition we plot the theoretically computed bispectrum. Firstly we note that the simulated initial conditions are in excellent agreement with the theoretical bispectrum, validating that our initial fields do contain the intended bispectrum shape. Note the deviations seen in the squeezed limit of the equilateral shape are statistical fluctuations that reduce when more simulations are considered. They are most visible in the squeezed limit of the equilateral shape as there the signal is smallest. Second it can clearly be seen that there are strong differences between the \emph{orthogonal-LSS} and \emph{orthogonal-CMB} templates. The differences are particularly large in the squeezed limit, as is expected, and will be very important for accurate modelling of effects such as scale-dependent bias \citep{Dalal_2008,Desjacques_2009}.

It is interesting to compare the primordial measurements to those at late times; in Fig. \ref{fig:pk_z0} we show the matter power spectrum derivative with respect to PNG at $z=0.0$. We see that the impact of PNG on the power spectrum for the \emph{local} and \emph{equilateral} shapes is similar in structure. Our results for \emph{local} and \emph{equilateral} are consistent with the results found in \citet{Wagner_2010}, which is a non-trivial check as \citet{Wagner_2010} use a different approach to generating the \emph{equilateral} shape initial conditions. We also show the 1-loop power spectrum prediction 
 from EFT of LSS \citep{DAmico_2022,Cabass_2022a,Cabass_2022b} - as expected we find excellent agreement on the largest scales and significant differences on small, non-perturbative scales, where the 1-loop term is inaccurate. Note that a subtlety of the theory model is that the counter term, required for the EFT of LSS to be a consistent theory, has an implicit dependence on $f_\mathrm{NL}$ which needs to be accounted for. The level of agreement is consistent with past work, e.g. \citet{Wagner_2010} who compared simulated power spectra with PNG at $z=0$ to the time renormalization group prediction. 

Finally in Fig. \ref{fig:bi_z0} we explore how the $z=0.0$ bispectrum is affected by PNG. We can see that at $z=0.0$ the tree level prediction for the bispectrum agrees on the largest scales, providing a simple validation of our simulations. However as we move towards smaller scales we find significant deviations from the tree-level prediction and the simulations --- as is expected. Our results are in agreement with previous investigations of the matter bispectrum in \citet{Sefusatti_2010}, who measured the bispectrum in simulations with \emph{local} non-Gaussianity up to $k_\mathrm{max}\approx 0.3~h/{\rm Mpc}$. Whilst our simulations are based off the same codes this check is an important validation of the robustness to choices of the simulation parameters, e.g. resolution and force softening. 
Our results for the local case are also in agreement with the results in \citet{Enrquez_2022}, who investigated the similarities between \emph{local} PNG and relativistic effects on the matter bispectrum. Note that the tree level approximation is the leading method for most analyses and forecasts \citep{Karagiannis_2018,Slosar_2019,Karagiannis_2020,Cabass_2022a,Cabass_2022b} to date. 1-loop perturbative methods \citep[e.g][]{Sefusatti_2009} have very recently been applied to galaxy survey measurements \citep{DAmico_2022} and have been used to extend range of scales that can be accurately modeled. These extract significantly more information about the primordial universe than the tree level calculations, but it is impossible to use such methods on the maximum scales considered here as they are non-perturbative.  
In this work we investigate whether we can utilize the  visible small scale information. 

\section{Fisher Methodology}\label{sec:FisherMethods}

\begin{figure}
    \centering
 \includegraphics[width=0.49\textwidth]{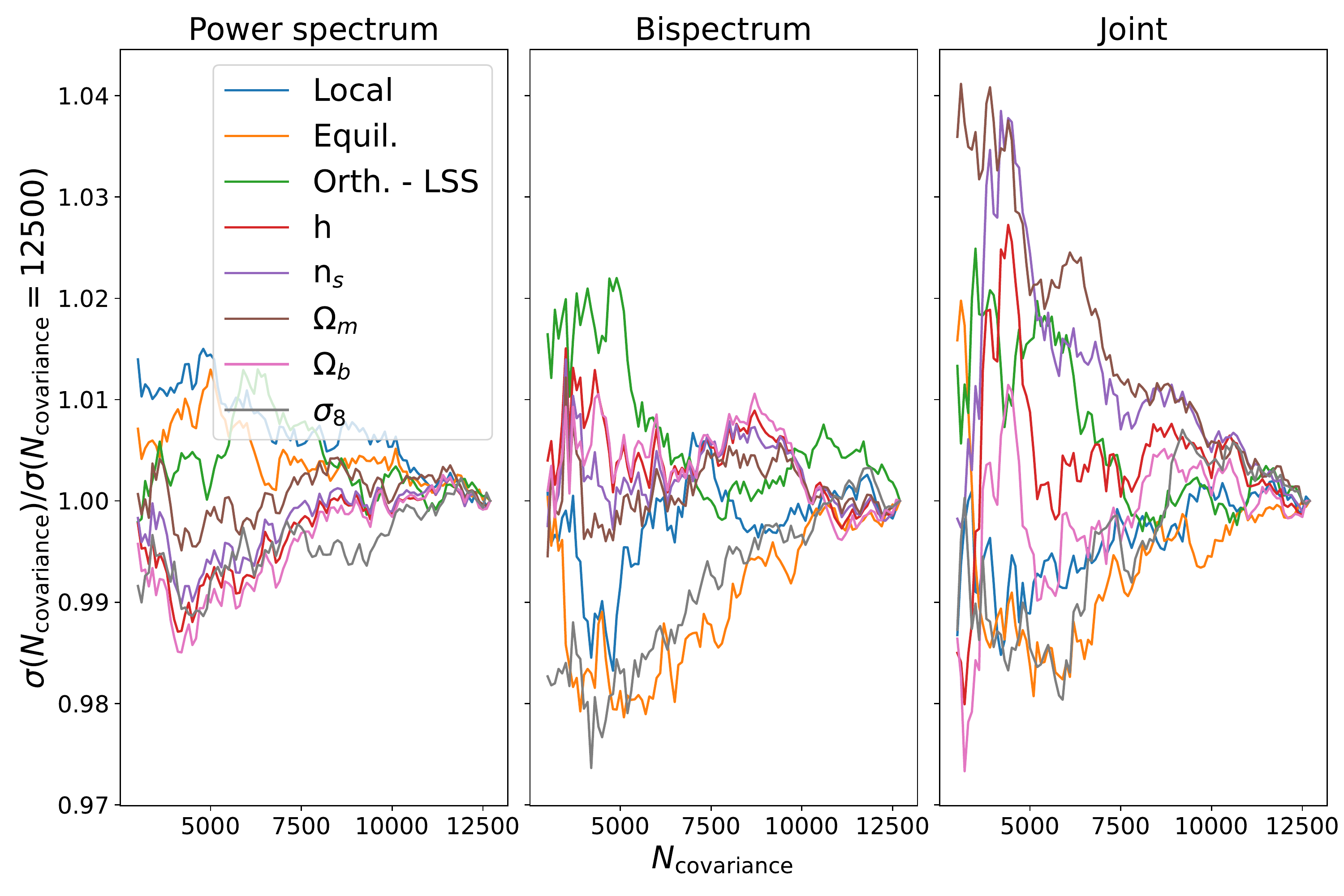}
 \caption{Variation in the marginalized parameter constraints as a function of number simulations used to compute the covariance matrix. The small fractional change suggests that the results are stable to the number of covariance simulations] \label{fig:convergence_test_covMat}}
\end{figure}
\begin{figure}
    \centering
     \includegraphics[width=0.49\textwidth]{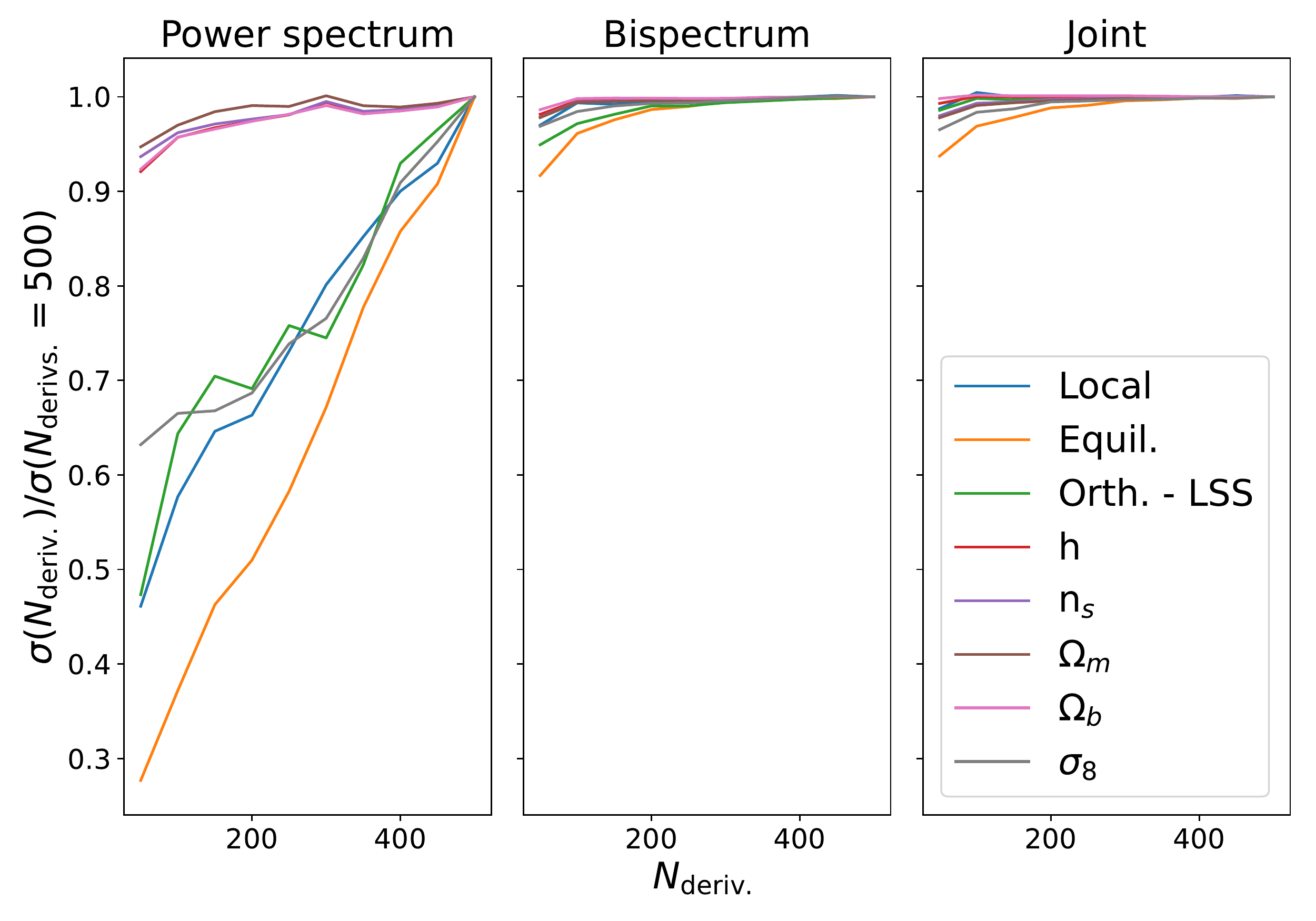}
     \caption{Variation in the marginalized parameter constraints as function of number simulations used to compute the Fisher derivatives. Whilst the bispectrum and joint statistics are converged, the power spectrum shows large changes and is likely unconverged. See the discussion in Section \ref{sec:FisherMethods}. \label{fig:convergence_test_derivs}}
\end{figure}
To estimate the information contained in the power spectrum and bispectrum we use the Fisher formalism \citep{Fisher_1935,Tegmark_1997}. The Fisher information, $F_{IJ}$, on parameters, $\mathbf{\theta}$, is defined as the variance of the score
\begin{align}
   F_{IJ} = \left<  \frac{\partial\log\mathcal{L}(\mathbf{X}|\mathbf{\theta})}{\partial \theta_I} \frac{\partial\log\mathcal{L}(\mathbf{X}|\mathbf{\theta})}{\partial \theta_J} \right>,
\end{align}
where $\mathcal{L}(X|\mathbf{\theta})$ is the likelihood. The Fisher information is useful as the minimum variance of an unbiased estimator, $\hat{\mathbf{\theta}}$, for $\mathbf{\theta}$ is given by \citep{Frechet_1943,aitken_silverstone_1942,Darmois_1945}
\begin{align}
\mathrm{Var}[\hat{\mathbf{\theta}}] = F_{II}^{-1},
\end{align}
with no summation over the repeated indices.
Thus by computing the inverse of the Fisher information we can infer the maximum information that we can learn about $\theta$ from measurements of an observable, $\mathbf{X}$.

In this work we assume that our observables, the matter power spectrum and bispectrum, are well approximated by a normal distribution. Thus
\begin{align}
   &\log\mathcal{L}(\mathbf{X}|\mathbf{\theta}) = \nonumber \\ &-\frac{1}{2} \sum_{ij}\left( O(k_i)-\bar{O}(k_i)\right)\Sigma_{ij}^{-1}\left(O(k_j)-\bar{O}(k_j)\right).
\end{align} where $O(k_i)$ denotes either the power spectrum or bispectrum, $\bar{O}$ is the observable mean and $\Sigma$ is the covariance matrix. We also consider the information available from joint measurements of the power spectrum and bispectrum, in which case $O(k_i)$ is then the vector of both the power spectrum and bispectrum measurements.  Whilst the Gaussian likelihood assumption is not perfectly accurate \citep[see e.g.][]{Scoccimarro_2000,Sellentin_2018}, it is sufficiently accurate to get an estimate of the information available.

For a Gaussian distribution, whose covariance is independent of the parameters of interest, the Fisher Information can be rewritten as
\begin{align}
    F_{IJ} = \frac{\partial\bar{O}(k_i)}{\partial\theta_I}\Sigma_{ij}^{-1}\frac{\partial\bar{O}(k_j)}{\partial\theta_J}.
\end{align}
In practice the covariance matrix of the power spectrum and bispectrum will likely have some cosmological dependence, however as was shown in \citet{Carron_2013} neglecting this dependence gives a better approximation for the true information of statistics like the power spectrum, whose true distribution is close to but not exactly Gaussian.

Thus to assess the information content in the bispectrum and power spectrum we require two ingredients, the derivative of our statistics of interest with respect to the parameters and the covariance of the statistics.

The derivatives are computed numerically by central differencing: we compute mean power spectra and bispectra signal at $\theta+\delta \theta$ and $\theta-\delta\theta$ and estimate the derivative as
\begin{align}
    \widehat{\frac{\partial \bar{O}(k_i)}{\partial \theta_I}} \approx \frac{\bar{O}(k_i)|_{\theta=\theta+\delta\theta} - \bar{O}(k_i)|_{\theta=\theta-\delta\theta}}{2 \delta \theta}.
\end{align}
for the PNG derivatives we use our new simulations, which where run with $\delta f_{\mathrm{NL}}=\pm 100$ and for the other cosmological parameters considered ($\Omega_m$, $\Omega_\Lambda$, $\Omega_b$, $\sigma_8$, $h$) we use the simulations summarized in Table 1 in \citet{Villaescusa-Navarro_2020}. 

The covariance matrix is the second important part of our analysis. The contributions to the covariance matrix can be broken down into three terms
\begin{align} \label{eq:covmat_terms}
    \Sigma_{ij} = C^{\mathrm{Gaussian}}_{ij}+C^{\mathrm{connected}}+C^{\mathrm{SSC}}
\end{align}
the Gaussian contribution, $C^{\mathrm{Gaussian}}_{ij}$, is straightforward to compute, whereas the connected, i.e. non-Gaussian, $C^{\mathrm{connected}}$, and super sample covariance terms $C^{\mathrm{SSC}}$ are generally difficult to compute \citep[see e.g][]{Kayo_2013,Chan_2017,Gualdi_2018,Sugiyama_2020}. We compute the first two terms by measuring the covariance of the observables using 12500 simulations in the fiducial cosmology (given the highly converged covariance matrix, we did not need to use all 15000 simulation available in the \textsc{quijote} suite) . The super sample covariance term is computed as in \citet{Li_2014,Chan_2018}, the details are summarized in Appendix \ref{app:covarianceMatrix}. Our base analysis does not include the super sample covariance term, but in Section \ref{sec:covMat} we assess the relative importance of it compared to the other terms. When inverting purely numerically estimated covariance matrices, we include a correction, the Hartlap/Anderson factor, to unbias our estimate of the inverse covariance matrix, precision matrix \citep{Hartlap_2007}.

We validate the accuracy of our results by verifying that the derivatives and covariance matrix are converged. In Fig. \ref{fig:convergence_test_covMat} and Fig. \ref{fig:convergence_test_derivs} we examine how our Fisher forecasts change as we vary the number of simulations used to estimate the two components of our forecast. We see that the forecast is highly stable against variations in the number of simulations used in the calculation. We also see that the bispectrum and the combined power spectrum and bispectrum --- hereafter joint --- forecasts are also stable to changes in the number of simulations used to estimate the derivatives. However we see that constraints from the power spectrum are not converged. This means that constraints from these statistics are likely over optimistic, indicating that actual constraints derived from power spectrum measurements will be reduced compared to the results reported below.

Typically one expects the power spectrum to converge more rapidly than higher order statistics so the convergence results in Fig. \ref{fig:convergence_test_derivs} are at first surprising. 
In the case of primordial non-Gaussianity however, the power spectrum is only sensitive to PNG through 1-loop and higher order terms, whereas the bispectrum is sensitive at the tree level. 
The power spectrum is thus only weakly sensitive to variations in the PNG parameters. Further, as will be seen in the Section \ref{sec:cosmo_constraints}, the effect of PNG on the power spectrum is highly degenerate with other parameters.  The strong degeneracies mean that the Fisher information needs to be estimated to a very high precision to be stable. Whilst the inferences can be regularized, e.g. by the use of a prior as is explored in Appendix \ref{app:priorPk}, in the main text we prefer to present these unconverged forecasts. Alternatively we could consider reparameterizations, for example those of the form $\sigma_8 f_\mathrm{NL}^\alpha$, which characterize the information probed by the power spectrum. However we find that the bispectrum both measures different combinations and allows inferences on the physical parameters, e.g. $f_\mathrm{NL}$, thus we retain the original parameterization. It is straightforward to see that the unconverged forecast is biased towards being overly constraining, see e.g. \citet{Coulton_2022b}, and it is interesting to compare how broad even these optimistic power spectrum constraints are when compared to the bispectrum ones. Note, if we exclude the PNG parameters from the Fisher forecast then the power spectrum forecasts are highly stable to variations in the number of derivative simulations used.

\section{Cosmological Constraints}\label{sec:cosmo_constraints}
\begin{figure*}
    \centering
    \includegraphics[width=\linewidth]{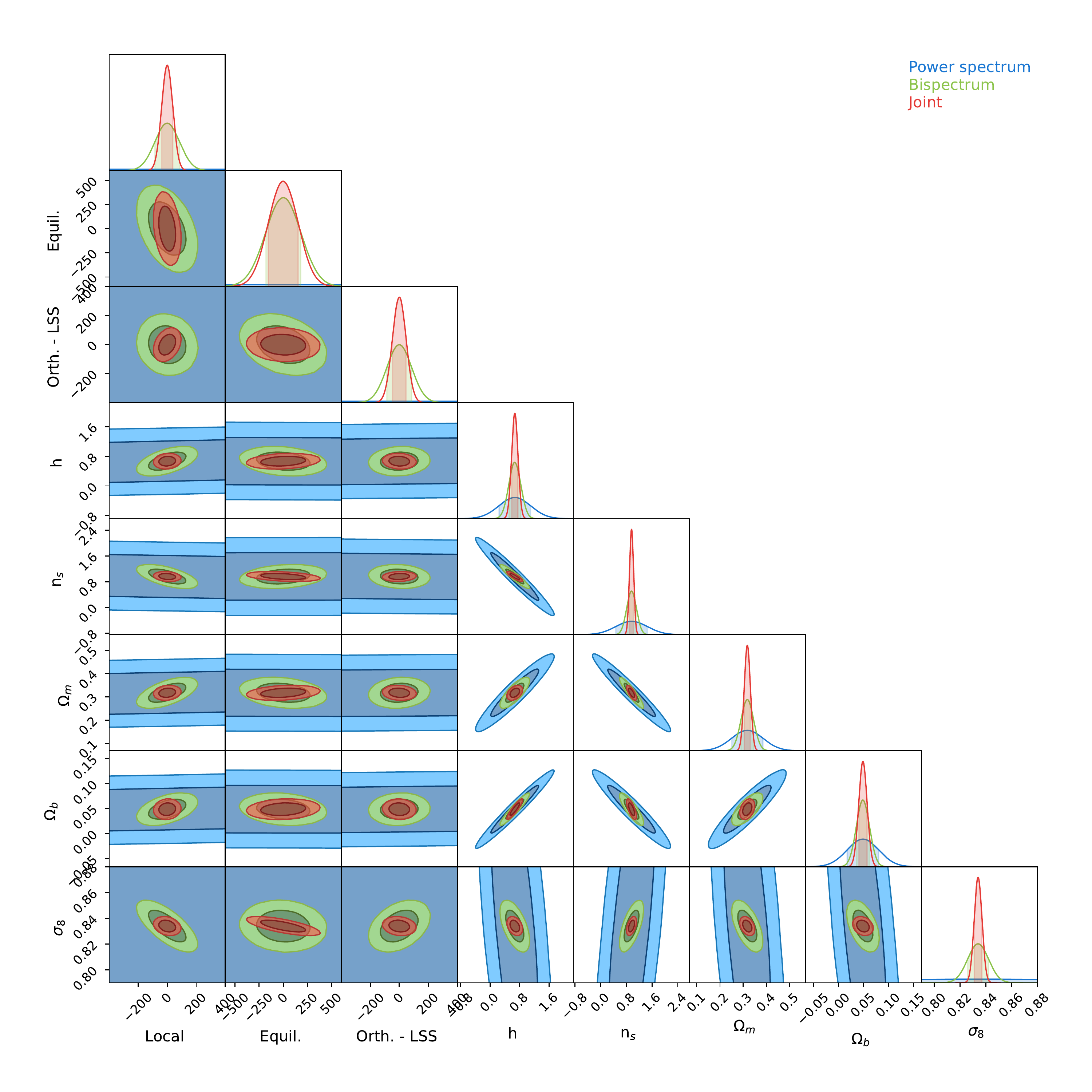}
    \caption{Joint constraints on the standard cosmological parameters and three shapes of primordial non-Gaussianity. \label{fig:param_const}}
\end{figure*}
Next we explore the constraining power from measurements at scales beyond the perturbation regime. Our constraints are obtained for a forecast experiment at $z=0.0$ and using modes from the fundamental mode, $k_{\rm F}=6.3\times10^{-3}~h/{\rm Mpc}$, to $k_\mathrm{max}=0.5~h/{\rm Mpc}$. 

\subsection{Degeneracies}\label{sec:param_degen}
\begin{figure*}
    \centering
    \includegraphics[width=\linewidth]{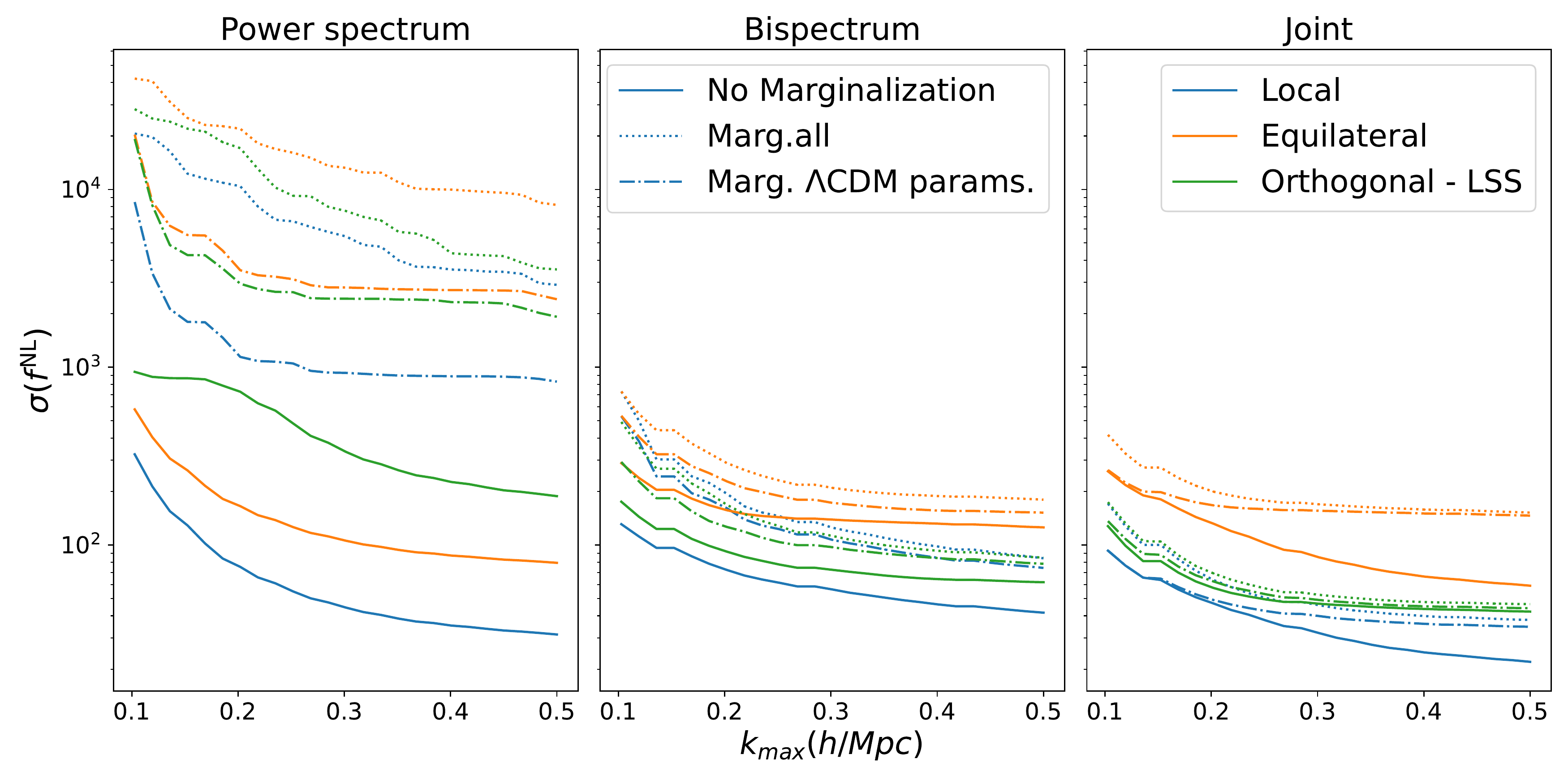}
    \caption{ The constraining power of the power spectrum, bispectrum and their combination as a function of scale for three shapes of non-Gaussianity. In solid lines we show the unmarginalized constraints, in dot-dashed we show the constraints when marginalizing over the cosmological parameters and in dotted we show the results from marginalizing over the cosmological parameters and the other shapes of primordial non-Gaussianity.  
    \label{fig:param_kmax}}
\end{figure*}

In Fig. \ref{fig:param_const} we plot the constraints obtainable with measurements of the matter power spectrum and bispectrum including modes up to $k_{\mathrm{max}}=0.5~h/{\rm Mpc}$. Focusing first on the constraints from only power spectrum measurements, we find that they have negligible information on primordial non-Gaussianity  when jointly measuring the different templates and marginalizing over cosmological parameters. This is expected given the small impact of PNG on the power spectrum and the similarity of the induced changes to other cosmological parameters (e.g. $n_s$ and $s_8$). Quantitatively the impact of marginalization, as seen in Fig. \ref{fig:param_kmax},  can widen constraints by up to a factor of $\sim 100$! Note that this marginalization also significantly degrades the cosmological parameter constraints - especially $\sigma_8$. The $\sigma_8 - f_\mathrm{NL}$ degeneracy is very strong as the PNG contributions to the matter power spectrum are very featureless, unlike for example $h$ which impacts the BAO, and can be largely captured by a rescaling of the amplitude. 

Focusing now on the bispectrum, we see that it offers vastly improved constraints. However, the bispectrum constraints also exhibit strong degeneracies with cosmological parameters. This indicates the challenge of separating the gravitationally sourced non-Gaussianity from PNG. When considering the joint power spectrum and bispectrum constraints, we see significant improvements that are largely driven by the improved constraining power on the standard, $\Lambda$CDM cosmological parameters. The joint analysis partially breaks the strong degeneracies (e.g. with $\Omega_m$ and $\sigma_8$) between the $\Lambda$CDM parameters,  leading to better PNG constraints. 

In Appendix \ref{app:extended}, we explore the degeneracies present when including two beyond $\Lambda$CDM parameters: $w$ and $\sum m_\nu$. 
\begin{figure}
    \centering
    \includegraphics[width=\linewidth]{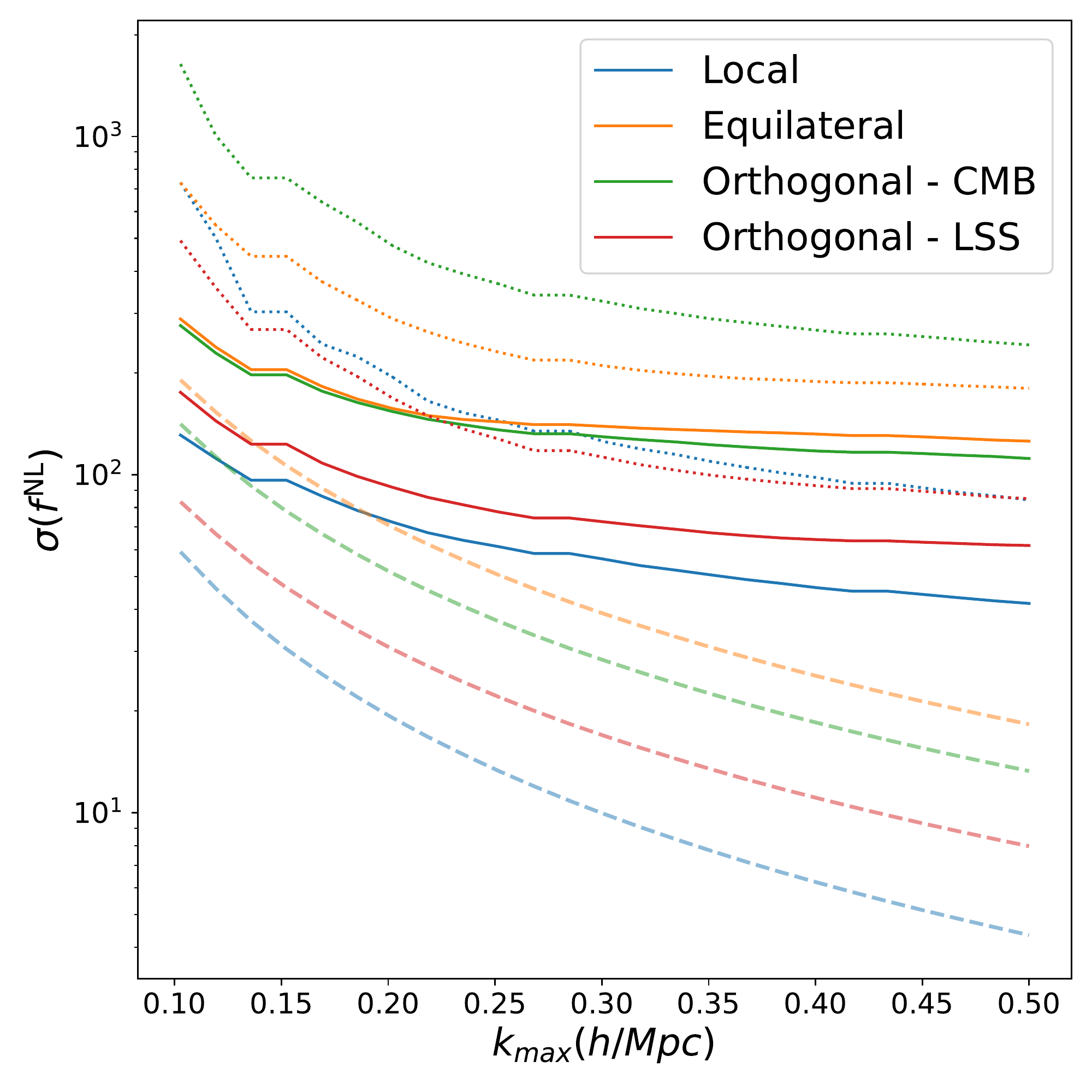}
    \caption{ The constraining power of the bispectrum as a function of scale compared to the information available in the primordial field. The dashed lines denote the constraining power in the primordial potential (without any marginalization), the solid lines denote the constraining power at $z=0.0$ without marginalization and the dotted lines are the constraints at $z=0.0$ with marginalization.  \label{fig:param_prim}}
\end{figure}
\begin{figure}
    \centering
    \includegraphics[width=\linewidth]{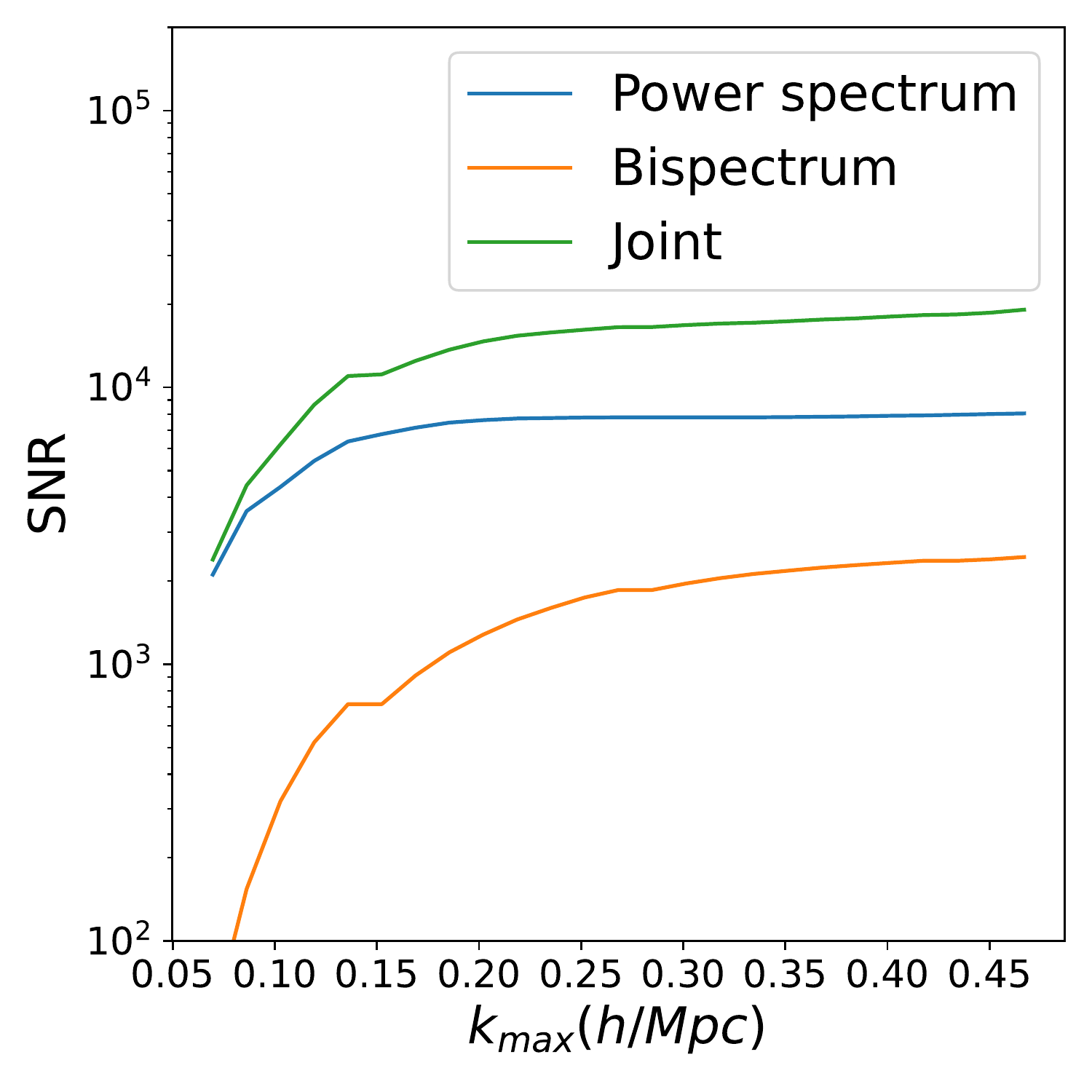}
    \caption{ The cumulative signal to noise (SNR) ratio for the power spectrum, bispectrum and joint analysis. \label{fig:SNR_cumulative}}
\end{figure}

\subsection{The Value of small scales}\label{sec:constraintsByScale}
Given the large information content available in the bispectrum, it is interesting to assess how much has been gained from the smaller scales. To answer this, we plot the constraining power of our probes as a function of maximum included scale in Fig. \ref{fig:param_kmax}. We perform this investigation for three cases 1) when constraining just one PNG shape, 2) when jointly constraining one PNG shape and the cosmological parameters and 3) when we jointly fit the three PNG shapes and the cosmological parameters. 

Focusing initially on case 1), we see that all probes yield significant gains by pushing the maximum scale up to $k_{\mathrm{max}}\approx 0.3~h/{\rm Mpc}$. However, the rate of improvement slows significantly beyond this scale. This arises due to the increasingly strong non-Gaussian contributions to covariance matrix, which we explore further in Section \ref{sec:covMat}. 
Importantly, the combined power spectrum and bispectrum analysis leads to significant improvements over analyses of either probe alone. This is explored more in \citet{Jung_2022}.

By considering cases 2) and 3), when we include the effects of marginalization, we can see the strength of the degeneracies. For the power spectrum the constraints degrade by up to two orders of magnitude.  Whilst the bispectrum constraints are less affected we still find degradation's of $\sim100\%$, reflecting the correlations seen earlier in Fig \ref{fig:param_const}. In this scenario there is more benefit to pushing to smaller scales as the extra information helps resolve the parameter degeneracies.  The joint constraint panel shows that combining the probes mitigates the impact of marginalization and again the constraints generally improve only modestly beyond $k_{\mathrm{max}}=0.3~h/{\rm Mpc}$. We also find minimal improvement for the \emph{equilateral} shape, further underlying the degeneracy between the primordial shape and gravitational bispectra.

To contextualize the expected improvement with scale,  we investigate the information available in the initial conditions. Given that the primordial universe is nearly Gaussian, we can compute the constraining power of an optimal estimator as
\begin{align}
 \sigma^{-2}= \frac{1}{6} \sum_{ijk}\frac{B^{\mathrm{theory}}(k_i,k_j,k_k)^2}{P_\Phi(k_i)P_\Phi(k_j)P_\Phi(k_k)}
\end{align}
where the sum counts the discrete Fourier modes in our simulation box. In Fig. \ref{fig:param_prim} we compare our constraints to the information available in the primordial bispectrum as a function of scale. As expected, the primordial constraints are tighter on all scales. Further, we see that pushing to increasingly small scales produces a far faster improvement on the constraining power \citep[see][ for a more detailed discussion]{Kalaja_2021}. As a cross check we have compared the primordial information captured if we were to use our sub-optimal, binned estimator. We find that the constraints degrade by $20-30\%$ . This could be reduced by optimizing the binning choice further (e.g. using a smaller bin width or having more bins for configurations where the bispectrum varies rapidly) or using an alternative estimator such as the modal estimator. 

In this analysis we also consider the information available to constrain the \emph{orthogonal-CMB} PNG but note that in our joint fits we only include either the \emph{orthogonal-CMB} or \emph{-LSS} shapes, not both. We do not include both simultaneously as they are both approximations to the same EFTi bispectrum and thus are expected to be highly degenerate. We see similar behavior in the constraints for the \emph{orthogonal-CMB bispectrum}.

\subsection{The importance of covariance matrix modelling}\label{sec:covMat}
The flattening of the information content arises as the bispectrum and power spectrum SNR flatten beyond $k \approx 0.3\ h/ {\rm Mpc}$ as is seen in Fig. \ref{fig:SNR_cumulative}. These results are consistent with those reported in \citet{Chan_2017}, who found that the bispectrum signal-to-noise ratio (SNR) is degraded by a factor of $\sim 10$ at $k_{\mathrm{max}}=0.5~h/{\rm Mpc}$ at $z=0.0$ and that the SNR flattens above $k_{\mathrm{max}}\approx0.3~h/{\rm Mpc}$.  This flattening arises as the nonlinear evolution of structure formation leads to strong correlations between the small scale modes, reducing the available primordial information. 

\begin{figure}
    \centering
    \includegraphics[width=\linewidth]{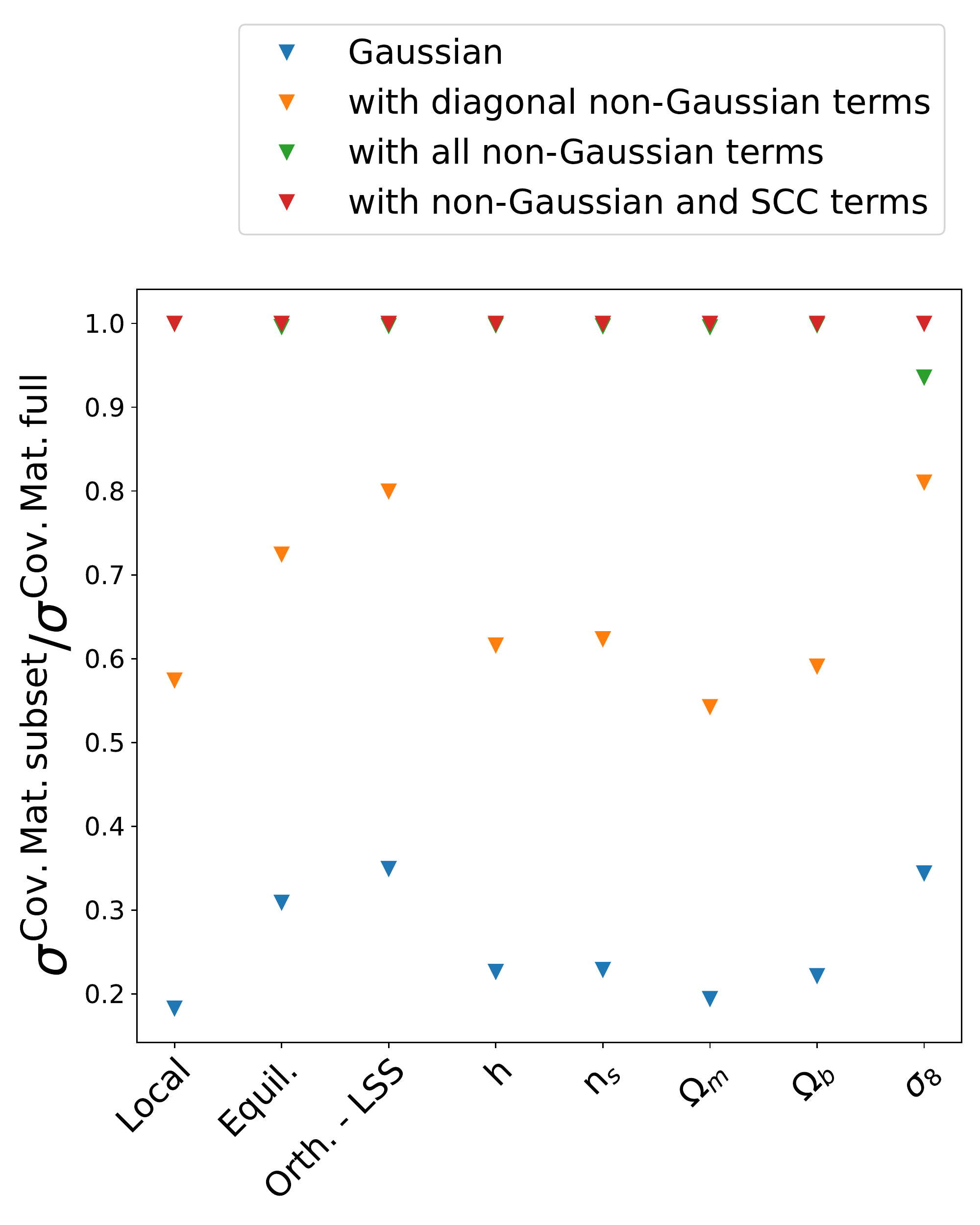}
    \caption{A comparison of the marginalized cosmological constraints from bispectrum-only  measurements when using only a subset of the contributions to the covariance matrix. The result is normalized by the errors obtained using all the terms listed in Eq. \ref{eq:covmat_terms}. \label{fig:covariance_comp}}
\end{figure}

To further understand the flattening of our constraints we explore how the different contributions to the covariance matrix propagate to parameter constraints. In Fig. \ref{fig:covariance_comp} we compared the bispectrum only constraints obtained with the full covariance matrix to the constraints obtained when only a subset of the contributions are included. 
We find that including the non-Gaussian contributions, both the diagonal and off-diagonal elements, has a significant impact on the constraining power and leads to a degradation of the constraints by a factor of $\sim 4$. This supports our assertion that the non-Gaussian terms cause the flattening of the SNR. These results agree with those seen in \citet{Barreira_2020,Biagetti_2021} and \citet{Gualdi_2020}; \citet{Biagetti_2021} find that the off-diagonal covariance matrix terms are highly important for constraints on \emph{local} PNG and, when included, lead to significantly degraded constraints. Our work extends these results highlighting their importance for other types of PNG.

Note that at smaller $k_\mathrm{max}$, the size of the non-Gaussian diagonal and off diagonal terms is reduced, but not negligible. At $k_\mathrm{max}=0.1~h/{\rm Mpc}$ there is still a degradation of some constraints by $\sim 2$.
It is interesting to note that the super sample covariance matrix terms have minimal impact on our parameter constraints despite being a non-trivial contribution to the covariance matrix as was found in \citet[e.g.][]{Hamilton_2006,Rimes_2006,Takahashi_2009,Chan_2018}. When we examine the unmarginalized constraints we find, for the power spectrum, that these super-sample covariance terms are important. This is similar to the results found in \citep{Li_2014b}. Thus for the power spectrum, the large increase in constraints due to marginalization swamps the impact of the super-sample covariance terms, at least for the setup considered here. For the bispectrum the limited degradation to the constraints arises as the strongly affected configurations are small-scale equilateral shapes \citep{Chan_2018,Barreira_2019}, which contribute minimally to the constraining power due to the large non-Gaussian covariance.

\section{Conclusions}\label{sec:conclusions}
In this work we have presented and validated a suite of N-body simulations that contain primordial non-Gaussianity: \textsc{quijote-png}. These augment the extensive \textsc{quijote} suite, matching the random seeds, configurations, and data formats of the previously run \textsc{quijote} simulations. This allows for consistent and seamless joint analyses of available information from the primordial universe and cosmological parameters. 

The simulation data, include particle data at $z=0.0,\,,0.5\,,1.0\,,2.0$ and $3.0$ and the power spectrum and bispectrum measurements at $z=0$ are publicly available.  Please refer to the documentation at \url{https://quijote-simulations.readthedocs.io/en/latest/png.html}

As a first use case, we investigated the information content accessible via measurements of the matter power spectrum and bispectrum. Our simulations allow us to extend the range of scales used in our analysis beyond the perturbative scales used in most advanced forecasts and analyses \citep[e.g.][]{Karagiannis_2018,Cabass_2022a,DAmico_2022}. This analysis, along with our companion paper \citep{Jung_2022}, are the first investigations of this kind. Our results show that significant gains can be made by including modes up to $\sim k_\mathrm{max} \approx 0.3~h/{\rm Mpc}$, after this the improvements are less dramatic. 
 
Beyond these scales, the large non-Gaussian covariance of our statistical probes limits improvements, despite large signals from PNG on these scales, by saturating the SNR of our probes. This saturation is not captured when Gaussian approximations are used to model the covariance matrix, thus highlighting the importance of accurately including the non-Gaussian terms. Similar saturation has been observed in the power spectrum \citep{Neyrinck2007,Rimes2005}, however these authors suggest that when pushing to even smaller, $k>1 \mathrm{h/Mpc}$, the information content again improves rapidly with scale. The resolution limitations of our simulations prevent us from investigating this, though we note that robustly exploiting such scales would be challenging to due the impact of baryonic processes \citep[see e.g.][]{Chisari2018}.  Interestingly, we find that on the scales considered here the super sample contributions to the covariance matrix are unimportant. In our companion paper \citet{Jung_2022} we examine the detailed features of the joint power spectrum - bispectrum covariance, which is an important component.

The challenge of modelling the large bispectrum covariance provides a large motivation to consider more efficient representations of the bispectrum information. Two such representation are explored in detail \citet{Jung_2022}. First we measure the modal bispectrum, which - by a judicious choice of basis functions- allows the information in the bispectrum to be fully captured in O(100) modal coefficients instead of the 1800 bispectrum bins used here. Second we then implement a further compressed estimator that compresses the information down to just 8 numbers. 

We chose to work at $z=0.0$, despite the limited volume of our Universe that is observable at $z=0.0$. This was done to allow a comparison to the information content in the halo field \citep{Coulton_2022a}, where resolution limitations mean we need to work at $z=0.0$ to obtain observationally relevant tracer densities. Measurements at higher redshifts will demonstrate similar features to those seen in our analysis with the modification that the non-linear scale will be moved to smaller scales - this is explicitly seen in our companion paper, \citet{Jung_2022}, where we investigate the information at $z=1.$. A second important choice is the size of box, and therefore $k_\mathrm{min}$, which is small compared to the volumes probed by upcoming experiments \citep[e.g.][]{dore_2013,DESI_2016}. PNG constraints are highly sensitive to the minimum scale used in the analysis, see  \citet{Kalaja_2021}  for an extended discussion. Thus the constraints on PNG obtainable from larger surveys will not just scale as $\sqrt{\mathrm{Vol.}}$ compared with those presented here. It is also expected that the degeneracies seen with cosmological parameters will be somewhat reduced as the shapes of the gravitationally induced and the primordial bispectra will be more easily differentiated over a larger domain. Running larger simulations, whilst maintaining the small scale resolution is computationally too expensive for this work, however our results demonstrate the importance of small scale measurements that will only be enhanced for larger volume surveys. 

A comparison between the information captured by our bispectrum measurements and the information available in the primordial field, which bounds how much we could possibly know about PNG, shows that the matter power spectrum and bispectrum estimators only capture $\sim 10\%$ of the total available information. This highlights the need to explore alternative approaches, such as topological measures \citep{Cole_2020}, the matter pdf \citep{Friedrich_2020}, machine learning approaches \citep{Giusarma_2019,Villaescusa-Navarro_2020} or field level approaches \citep{Andrews_2022}, to fully capture the information available in the matter field. The simulations presented here are designed to facilitate future investigations along these lines. 

Finally, it is important to emphasize that the 3D matter field considered in this work is not directly accessible. Observationally, we can access the 2D integrated matter field through gravitational lensing and these observations will be impacted by observational systematics and baryonic processes, which are neglected here. Both of these effects will likely impact the observationally attainable information. We can also make biased estimates of the matter field from surveys of biased tracers, such as galaxy positions and velocities, and signatures of the neutral hydrogen distribution through 21cm mapping. Thus we stress the results of this work are not forecast constraints for upcoming surveys, but provide an assesses the value of smaller scale bispectrum measurements and provide a suite of simulations for the community to use to develop and test new analysis tools for PNG. This work represents an important first step towards these more sophisticated probes. In a followup work we will explore more direct observables, such as the halo and galaxy fields.

\begin{acknowledgments}
The authors are very grateful to Oliver Philcox, Daan Meerburg, Simone Ferraro, Shirley Ho, David Spergel and Yin Li for their advice on numerous aspects of our analysis. 
GJ, ML and MB were supported by the project "Combining Cosmic Microwave Background and Large Scale Structure data: an Integrated Approach for Addressing Fundamental Questions in Cosmology", funded by the MIUR Progetti di Ricerca di Rilevante Interesse Nazionale (PRIN) Bando 2017 - grant 2017YJYZAH.
DK is supported by the South African Radio Astronomy Observatory (SARAO)
and the National Research Foundation (Grant No. 75415). B.D.W. acknowledges support by the ANR BIG4 project, grant ANR-16-CE23-0002 of the French Agence Nationale de la Recherche; and the Labex ILP (reference ANR-10-LABX-63) part of the Idex SUPER, and received financial state aid managed by the Agence Nationale de la Recherche, as part of the programme Investissements d'avenir under the reference ANR-11-IDEX-0004-02.
The Flatiron Institute is supported by the Simons Foundation.
LV acknowledges  ERC (BePreSySe, grant agree- ment 725327),  PGC2018-098866- B-I00 MCIN/AEI/10.13039/501100011033 y FEDER “Una manera de hacer Europa”, and the “Center of Excellence Maria de Maeztu 2020-2023” award to the ICCUB (CEX2019-000918-M funded by MCIN/AEI/10.13039/501100011033).
\end{acknowledgments}
\appendix
\begingroup
\renewcommand{\arraystretch}{1.4} 
\begin{table*}
    \centering
    \begin{tabular}{c c c c c}
         Bispectrum Term & Label($\alpha$) & $K_\alpha^{I}$ &$K_\alpha^{II}$ &$K_\alpha^{III}$   \\ \hline
        $\frac{1}{k_1^3k_2^3}$ & A & 1  & $\frac{1}{2} \frac{1}{k_3^3}\left[k_1^3+k_2^3 \right] $ &  - \\ 
        $\frac{1}{k_1k_2^2k_3^3}$ & B & $\frac{1}{k_3}\left[k_1+k_2\right]$ &   $\frac{1}{k_3^2}\left[k_1^2+k_2^2\right]$& $\frac{1}{k_3^3}\left[k_1^2k_2+k_1k_2^2\right]$   \\ 
        $\frac{1}{k_1^2k_2^2k_3^2}$ & C   & $\frac{1}{3}\frac{1}{k_3^2} k_1k_2$ & - & - \\ 
        $\frac{k_1^2}{k_2^4k_3^4}$ & D & $\frac{k_3^2}{k_1k_2}$  & $\frac{1}{2} \frac{1}{k_3^4}\left[\frac{k_1^5}{k_2}+\frac{k_2^5}{k_1} \right] $ & - \\ 
        $\frac{1}{k_1k_2k_3^4}$ & E & $\frac{k_1^2k_2^2}{k_3^4}$  & $\frac{1}{2} \frac{1}{k_3}\left[\frac{k_1^2}{k_2}+\frac{k_2^2}{k_1} \right] $ &  - \\ 
        $\frac{k_1}{k_2^3k_3^4}$ & F & $k_3\left[\frac{1}{k_1}+\frac{1}{k_2}\right]$ &  $\frac{1}{k_3^3}\left[\frac{k_1^4}{k_2}+\frac{k_2^4}{k_1}\right]$&   $\frac{1}{k_3^4}\left[k_1^4+{k_2^4}\right]$ \\ 
        $\frac{1}{k_2^2k_3^4}$ & G & $ \left[\frac{k_2}{k_1}+\frac{k_1}{k_2}\right]$ &  $\frac{1}{k_3^2}\left[\frac{k_1^3}{k_2}+\frac{k_2^3}{k_1}\right]$&   $\frac{1}{k_3^4}\left[k_1^3k_2+k_1k_2^3\right]$ \\ 
    \end{tabular}
    \caption{The kernels associated with each term in the bispectrum (as shown in the first column). Each bispectrum term can typically be generated by multiple kernels. Whilst the result is equivalent at the bispectrum level, for other statistics they have different properties. \label{tab:kernels}}
   
\end{table*}
\endgroup
\section{Initial condition generation}\label{app:ic_kernels}
The kernels used to generate initial conditions with \emph{local}, \emph{equilateral} and \emph{orthogonal-CMB} non-Gaussianity are identical to those used in \citet{Scoccimarro_2012}. We use the notation
\begin{align}
  &  K^\alpha_A =\int \frac{\mathrm{d}^3 k_1}{(2\pi)^3}\frac{\mathrm{d}^3 k_2}{(2\pi)^3} K^\alpha_A(\mathbf{k}_1,\mathbf{k}_2,\mathbf{k}) \Phi(\mathbf{k_1})\Phi(\mathbf{k_2})\times \nonumber   \\ &(2\pi)^3\delta^{(3)}(\mathbf{k}_1+\mathbf{k}_2+\mathbf{k})
\end{align}
and the set of kernels defined in Table \ref{tab:kernels}
As in   \citet{Scoccimarro_2012}, we generate \emph{local} non-Gaussianity using
\begin{align}
    \Phi^{\mathrm{local}}= \Phi+f_\mathrm{NL}^{Local}K_A^I
\end{align}
\emph{equilateral} with
\begin{align}
    \Phi^{\mathrm{Equil.}}= \Phi+f_\mathrm{NL}^{Equil}\Big[ -3K_A^I-6K_C^I+4K_B^I+2K_B^{II}\Big]
\end{align}
and \emph{orthogonal-CMB} with
\begin{align}
    \Phi&^{\mathrm{Ortho-CMB}}= \Phi  \nonumber \\ & +f_\mathrm{NL}^{Ortho-CMB}\Big[ -9K_A^I-24K_C^I+10K_B^I+8K_B^{II}\Big]
\end{align}

\subsection{Orthogonal-LSS}
 In this section we derive the kernels used in Eq. \ref{eq:kernelConv} to generate initial conditions with the \emph{orthogonal-LSS} non-Gaussianity. This derivation mirrors the derivation and choices made for the other shapes in  \citet{Scoccimarro_2012}. 
The bispectrum in Eq. \ref{eq:bis_or_lss} is separable and the first three terms are of the same form as the \emph{equilateral} and f terms used in \citet{Scoccimarro_2012}. The next step is to enumerate the possible kernels associated with each term and these are listed in Table \ref{tab:kernels}.

Using this we can show that the \emph{orthogonal-LSS} bispectrum can be generated by the following operations
\begin{align}
&\frac{\Phi^{NG}(\mathbf{k})}{3 } \nonumber \\ &= -\left[1+\frac{9p}{27} \right]\Big[(1-u)K^I_A+\frac{u}{2}K^{II}_A \Big]  -\left[2+\frac{60p}{27} \right]\frac{1}{3}K_C^I \nonumber \\
&+\left[1-\frac{15p}{27} \right]\Big[ (1-t-s)K^I_B+tK^{II}_B+sK^{III}_C\Big] \nonumber \\
&+\frac{p}{27}\Big[(1-a) K_D^I+\frac{a}{2}K_D^{II} \Big]  \nonumber \\
&-\frac{20p}{27}\Big[(1-b) K_E^I+\frac{b}{2}K_E^{II} \Big]  \nonumber \\
&+\frac{6p}{27}\Big[(1-c-d) K_F^I+cK_F^{II}+dK_F^{III} \Big]  \nonumber \\
&+\frac{15p}{27}\Big[(1-e-f) K_G^I+eK_G^{II}+fK_G^{III} \Big].
\end{align}
There is a large family of fields with the correct bispectrum with differing $N\ne 3$-point functions. Our approach to choosing the terms is to impose the constraint that the largest corrections to the power spectrum should scale as $k^{-2}$ and, for simplicity, to set as many of the remaining coefficients as possible to zero. For two kernels $K^A(k_1,k_2,k_3) = k_1^ak_2^bk_3^c$ and $K^B(k_1,k_2,K_3)=k_1^\alpha k_2^\beta k_3^\gamma$ the leading correction to the power spectrum is given as
\begin{align}
    \delta P(k_3) &= k_3^{c+\gamma}\int\frac{\mathrm{d}^3k_1}{(2\pi)^3}\Bigg[k_1^{a+\alpha-3}(|\mathbf{k_1}+\mathbf{k_3}|)^{b+\beta-3} \nonumber \\ & + k_1^{b+\alpha-3}(|\mathbf{k_1}+\mathbf{k_3}|)^{a+\beta-3} \Bigg]. 
\end{align}
Thus to satisfy our constraints we require
\begin{align}
    \frac{15p}{27}f-\frac{6p}{27}d - \frac{20p}{27}\frac{(1-b)}{2}+\frac{p}{27}\frac{a}{2} = 0 \nonumber \\
    -\left(1+\frac{9p}{27} \right)\frac{u}{2}+ \left(1+\frac{15p}{27}\right) s -\frac{6p}{27} c = 0 \nonumber \\
    \left(1+\frac{15p}{27} \right) t +\frac{15p}{27}e -\frac{1}{6}\left(2+\frac{60p}{27}\right) = 0.
\end{align}
We thus choose
\begin{align}
    b = 1 \nonumber \\
    t =  \frac{\left(2+\frac{60p}{27}\right)}{6 \left( 1 + \frac{15p}{27} \right)} \\
    a=c=d=e=f=u=s=0.
\end{align}
Bringing these terms together we have
\begin{align}
&\frac{\Phi^{NG}(\mathbf{k})}{3 }= -\left[1+\frac{9p}{27} \right]K^I_A  -\left[2+\frac{60p}{27} \right]K_C^I \nonumber \\
&+\left[1+\frac{15p}{27} \right]\Big[ (1-t)K^I_B+tK^{II}_B\Big] \nonumber \\
&+\frac{p}{27}K_D^I-\frac{20p}{27}K_E^{II}+\frac{6p}{27}K_F^I+\frac{15p}{27}K_G^I
\end{align}
or equivalently
\begin{align}
\frac{\Phi^{NG}(\mathbf{x})}{3 }&=
 -\left[1+\frac{9p}{27} \right]\Phi^2 -\left[2+\frac{60p}{27} \right]\frac{1}{3}\partial^{-2}\left[ \partial\Phi\partial\Phi \right]\nonumber \\
&+2\left[1+\frac{15p}{27} \right]\Big[ (1-t)\partial^{-1}\left[\Phi\partial\Phi\right]+t\partial^{-2}\left[\Phi\partial^2\Phi\right]\Big] \nonumber \\
&+\frac{p}{27}\partial^2\left[\partial^{-1}\Phi\partial^{-1}\Phi\right]-\frac{20p}{27}\partial^{-1}\left[\partial^2\Phi\partial^{-1}\Phi\right] \nonumber \\ 
&-\frac{12p}{27}\partial\left[ \Phi\partial^{-1}\Phi\right]+\frac{30p}{27}\partial\Phi\partial^{-1}\Phi.
\end{align}
These operators have been derived assuming a scale invariant spectrum ($n_s=1$) for simplicity. It is simple to adapt to a non-scale invariant spectrum by replacing  $k$ with $P(k)^{-\frac{1}{3}}$. Whilst this replacement alters the primordial bispectrum, the distortion is small, as $n_s\sim 1$, and it allows us to simply use non-scale invariant simulations in our analysis (i.e. $n_s\ne 1$) as is required to be consistent with observations \citep{planck2016-l06}.

\section{Covariance matrix computation}\label{app:covarianceMatrix}
The covariance matrix given in Eq. \ref{eq:covmat_terms} is composed of several terms, the Gaussian, connected and super sample covariance terms. 
\subsection{Gaussian Covariance matrix}
The computation of the Gaussian covariance matrix is for the power spectrum estimator, Eq. \ref{eq:ps_estimator}, is given by \citep{Feldman_1994}
\begin{align}
    C^{\mathrm{Gaussian}}_{ij} = \frac{2}{N^2_{\mathrm{modes}}}\sum\limits_{k_\alpha \in k_i} P(k_\alpha)^2 \delta_{k_i,k_j}
\end{align}
where $N_{\mathrm{modes}}$ is the number of modes in the power spectrum binned.  For the binned bispectrum estimator the equivalent expression is
\begin{align}
   C^{\mathrm{Gaussian}}&(\hat{B}_{ijk},\hat{B}_{IJK}) \nonumber \\ & =\frac{g^{IJK}_{ijk}}{N^2_{\mathrm{triplets}}} \sum P(k_\alpha)P(k_\beta)P(k_\gamma)
\end{align}
where $N_{\mathrm{triplets}}$ is the number of bispectrum configurations in the bin, the sum is over all bispectrum triplets and $g^{IJK}_{ijk}$ selects only the diagonal elements of the covariance matrix and is 1,2 or 6 depending on whether there are 3,2 or 1 unique values in $\{ijk \}$.
\subsection{Super sample covariance}\label{sec:super_sample_cov}
Generally the super-sample covariance contribution to the covariance matrix is given by
\begin{align}\label{eq:ssc_term}
    C^{SSV}_{X,Y}(k,k') = \sigma^2_W \left.\frac{\partial X(K)}{\partial \delta_b}\right|_{\delta_b=0} \left.\frac{\partial Y(K')}{\partial \delta_b}\right|_{\delta_b=0}
\end{align}
where
\begin{align}
    \sigma^2_b = \int \frac{\mathrm{d}^3k}{(2\pi)^3} \frac{W(\mathbf{k})}{V_W}\frac{W(\mathbf{k})}{V_W} P_L(k).
\end{align}
We use follow \citet{Li_2014} and use separate universe simulations to compute the responses to a long wavelength mode (the derivative terms in Eq. \ref{eq:ssc_term}).
For the power spectrum of global mean observables this is given by
\begin{align}
   \left.\frac{\partial P(K)}{\partial \delta_b}\right|_{\delta_b=0}  = P(k) + \frac{\partial P_{SU}(K)}{\partial \delta_b} -\frac{1}{3}\frac{\partial P(k)}{\partial \ln k},
\end{align}
and for local power spectra observables we have an additional $-2P(k)$.
For the bispectrum of global mean observables this is  \citep{Chan_2018}
\begin{align}
    \left.\frac{\partial B(k_1,k_2,k_3)}{\partial \delta_b}\right|_{\delta_b=0}  =& B(k_1,k_2,k_3) + \frac{\partial B_{SU}(k_1,k_2,k_3)}{\partial \delta_b} \nonumber \\ & -\sum\limits_i\frac{1}{3}\frac{\partial B(k_1,k_2,k_3)}{\partial \ln k_i}
\end{align}
and for local mean observables there is an additional $-3 B(k_1,k_2,k_3)$. In both case the derivative $\partial P_{SU}(k)/\partial \delta_b$ and $\partial B_{SU}(k_1,k_2,k_3)/\partial \delta_b$ is computed from the separate universe simulations using finite difference.

\begin{figure}
    \centering
    \includegraphics[width=\linewidth]{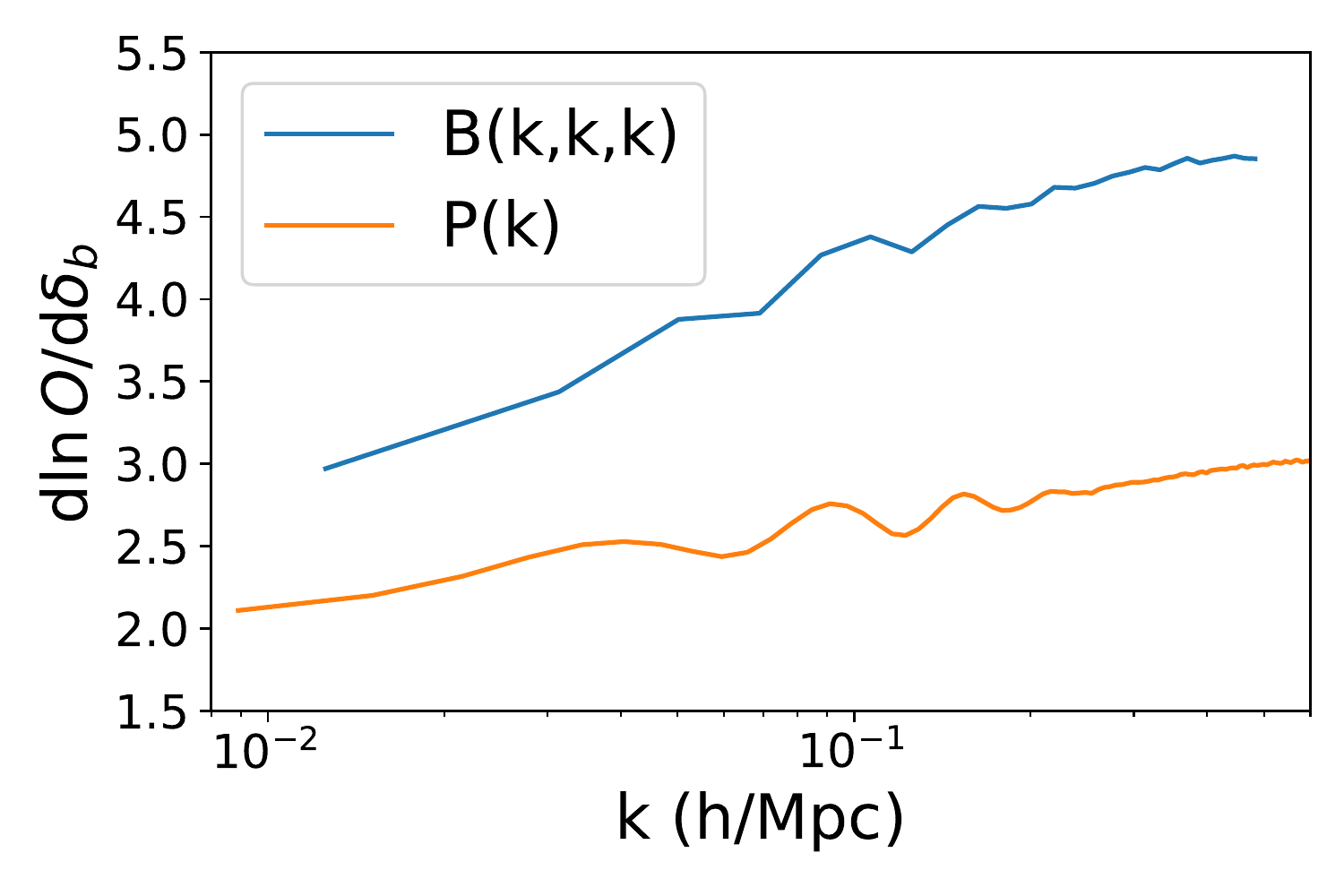}
    \caption{The response functions of the power spectrum and bispectrum to a long wavelength mode. These have been computed using separate universe simulations as described in \citet{Li_2014}  \label{fig:ssc_responses}}
\end{figure}

In Fig. \ref{fig:ssc_responses} we plot the response functions and these show great agreement with the equivalent plots in \citet{Li_2014} and \citet{Chan_2018}. Further we find that the relative importance of the super-sample covariance terms is very similar to that found in \citet{Chan_2018} providing validation of our implementation.

\begin{figure*}
    \centering
    \includegraphics[width=\linewidth]{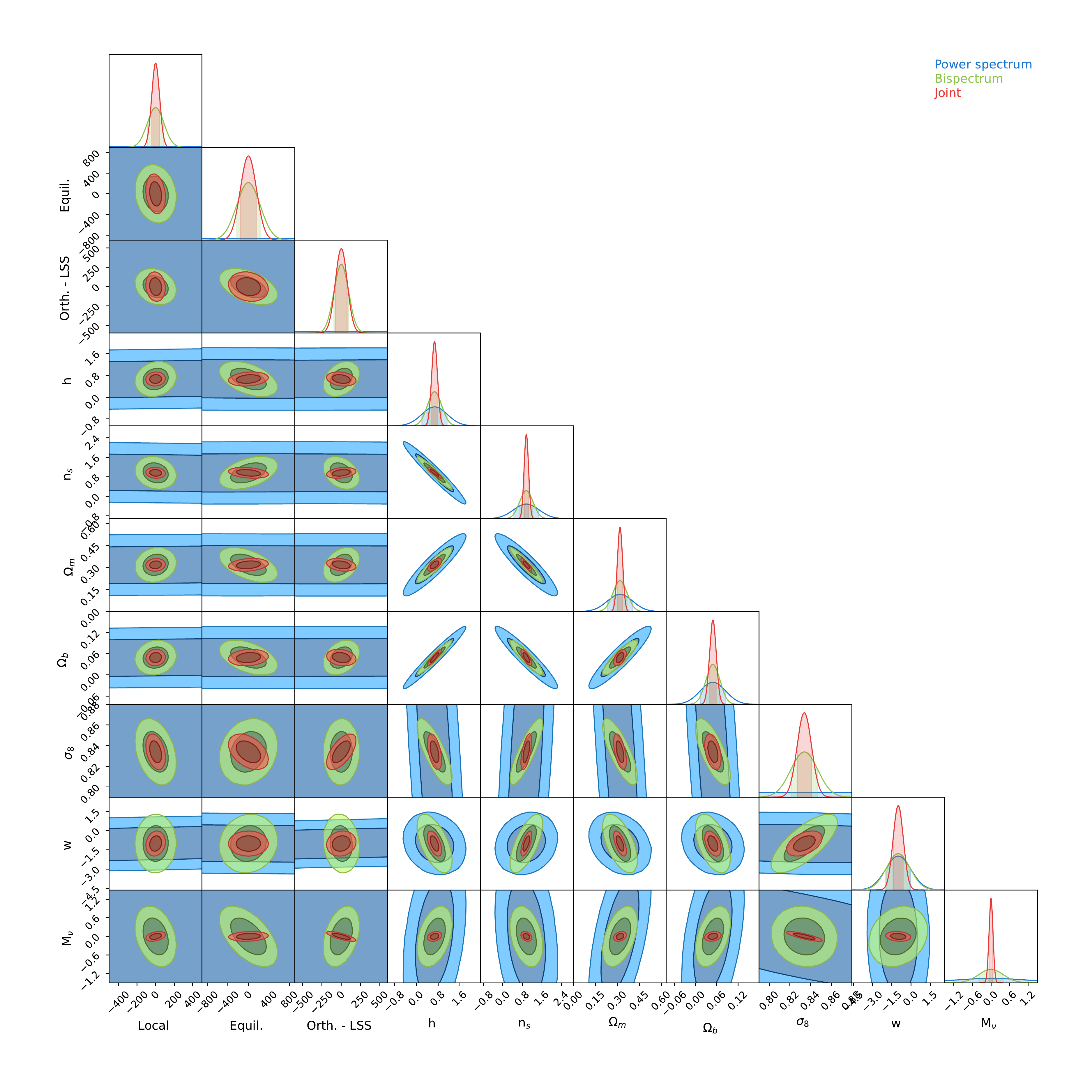}
    \caption{Joint constraints on the standard cosmological parameters, the equation of state of dark energy, $w$, the sum of the masses of the neutrinos, $\sum m_\nu$, and three shapes of primordial non-Gaussianity. \label{fig:param_const_wMnu} }
\end{figure*}
\begin{figure}
    \centering
    \includegraphics[width=\linewidth]{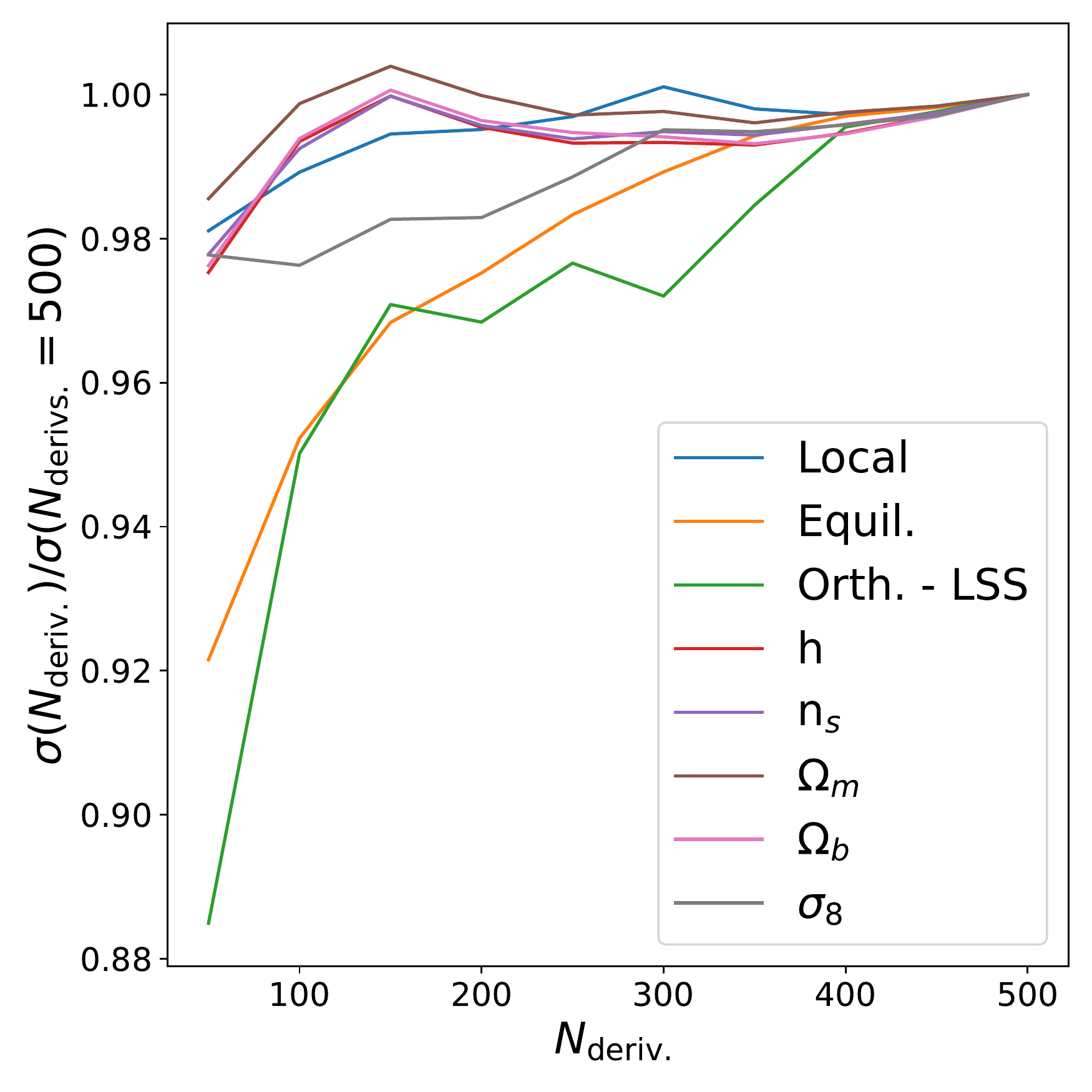}
    \caption{The convergence of the power spectrum measurements when using a Gaussian prior of width $f_\mathrm{NL}=1000$. \label{fig:convergenceWithPrior}}
\end{figure}

\begin{figure*}
    \centering
    \includegraphics[width=\linewidth]{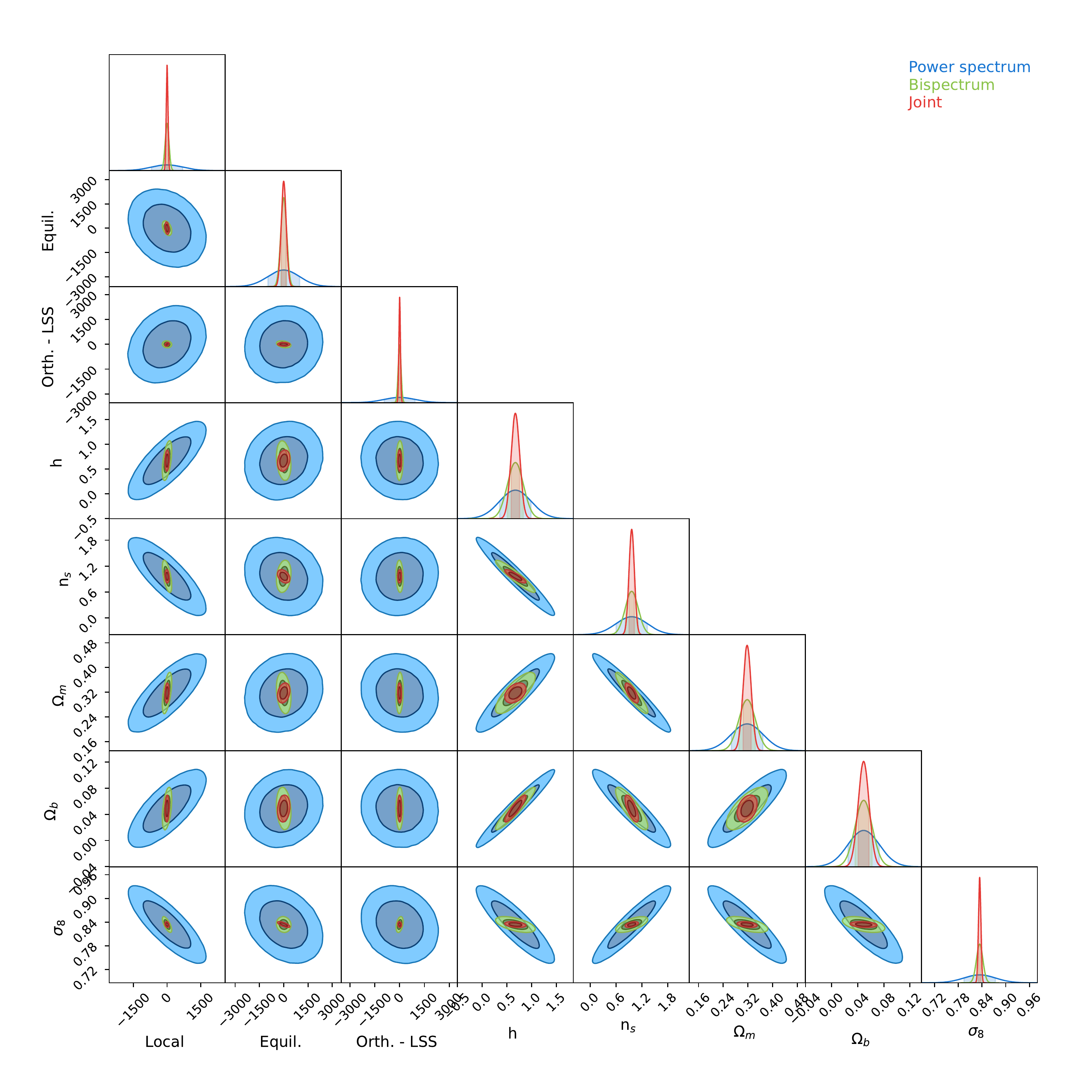}
    \caption{The cosmological constraints obtained when a Gaussian prior of width $\sigma(f_\mathrm{NL})=1000$ is applied to the power spectrum. \label{fig:param_const_wPrior}}
\end{figure*}
\section{Joint constraints with beyond $\Lambda$CDM parameters}\label{app:extended}
 In this Appendix we extend our analysis to include two common extensions: the dark energy equation of state parameter, $w$, and the sum of the masses of the neutrinos $\sum m_\nu$. We use the simulations described in \citet{Villaescusa-Navarro_2020} to compute the derivatives with respect to $w$ and the simulations in \citet{Hahn_2020} for the neutrino mass constraints. For the neutrino mass we use two modifications to our method: firstly we compute the properties with respect to the total matter field (rather than just the dark matter field). This is trivial done by include the neutrinos when gridding the simulation outputs. Second we use the third order difference method, Eq. 4.5 in  \citet{Hahn_2020}, to compute the derivatives. This is used as the neutrino mass is positive definite and thus we cannot use the central difference method used for the other parameters. Additionally this higher order method provides a more accurate estimate of the derivatives.
 
 In Fig. \ref{fig:param_const_wMnu}, we plot the joint constraints of all the cosmological and PNG parameters. The inclusion of $w$ primarily degrades the constraints on PNG by decreasing the constraining power on the $\Lambda$CDM parameters. On the other hand we find strong degeneracies between $\sum m_\nu$ and the $f_\mathrm{NL}$ parameters. This is not unexpected given that the PNG parameters exhibit degeneracies with $\sigma_8$ and the impact of $ \sum m_\nu$ is similar to $\sigma_8$ on these scales.

\section{Impact of a prior on the power spectrum convergence}\label{app:priorPk}
In Section \ref{sec:FisherMethods} we found that the power spectrum constraints were not converged. As is seen in Fig \ref{fig:param_kmax} the power spectrum constraints are increased by a factor of $\sim 10-100$ when marginalizing and this large degeneracy is the reason for the lack of convergence. To test this idea we perform our convergence test when imposing a Gaussian prior of width $f_\mathrm{NL}=1000$ for all the PNG shapes. In Fig. \ref{fig:convergenceWithPrior} we see the convergence in this case finding that all derivatives are sufficiently converged. We show the parameter constraints obtained when including this prior in Fig. \ref{fig:param_const_wPrior}. The only effect of the prior is for the power spectrum constraints where it improves the constraints on almost all of the cosmological parameters by $\sim 20\%$ with the exception of $s_8$ which is significantly improved. 

\bibliographystyle{apsrev.bst}
\bibliography{png,Planck_bib}

\end{document}